\documentclass[12pt]{article}

\usepackage{amssymb,amsmath,amsfonts,eurosym,geometry,ulem,graphicx,caption,color,setspace,sectsty,comment,footmisc,natbib,pdflscape,subcaption,array,hyperref,threeparttable,todonotes,float,xr-hyper,multirow,rotating,booktabs,placeins,indentfirst,adjustbox,tabularx,bm,makecell}
\setcitestyle{authoryear,round}
\usepackage{xurl}

\usepackage{etoolbox}
\AtBeginEnvironment{table}{\footnotesize}
\newcommand{\mA}{\mathcal{A}}

\newcommand{\mC}{\mathcal{C}}

\newcommand{\mG}{\mathcal{G}}

\newcommand{\mJ}{\mathcal{J}}

\newcommand{\mM}{\mathcal{M}}

\newcommand{\mR}{\mathcal{R}}

\newcommand{\bE}{\mathbb{E}}

\newcommand{\bR}{\mathbb{R}}

\newcolumntype{L}[1]{>{\raggedright\let\newline\\arraybackslash\hspace{0pt}}m{#1}}
\newcolumntype{C}[1]{>{\centering\let\newline\\arraybackslash\hspace{0pt}}m{#1}}
\newcolumntype{R}[1]{>{\raggedleft\let\newline\\arraybackslash\hspace{0pt}}m{#1}}

\hypersetup{colorlinks=true, linkcolor=black, citecolor=blue, urlcolor=blue}

\begin{document}

\begin{titlepage}
\title{Carbon Regulation and Competition in the \\ European Airline Industry\thanks{We are grateful to Andreas Sch\"afer, Lynette Dray, and Khan Doyme from the UCL Air Transportation Systems Lab for sharing data and insights on the European airline industry. We especially thank Christian Bontemps and participants at the 2026 Symposium on Aviation Research for valuable comments and discussion. We gratefully acknowledge the use of the UCL Myriad High Performance Computing Facility (Myriad@UCL) and associated support services, and the Isambard 3 Tier-2 HPC Facility hosted by the University of Bristol and operated by the GW4 Alliance, funded by EPSRC [EP/X039137/1]. All errors are our own.}}

\author{Ertian Chen\thanks{School of Economics \& Paula and Gregory Chow Institute for Studies in Economics, Xiamen University. Email: \href{mailto:ertianchen@xmu.edu.cn}{ertianchen@xmu.edu.cn}.}
\and
Lichao Chen\thanks{Cambridge Econometrics. Email: \href{mailto:lichaochen331@gmail.com}{lichaochen331@gmail.com}.}
\and
Lars Nesheim\thanks{University College London, Bristol, CeMMAP and IFS. Email: \href{mailto:l.nesheim@ucl.ac.uk}{l.nesheim@ucl.ac.uk}.}
}
\date{October 18, 2026}
\maketitle

\vspace{-2cm}
\begin{center}
\end{center}

\begin{abstract}
\noindent The European Union Emissions Trading System is set to substantially increase the effective carbon price faced by airlines. We estimate a two-stage model of airline competition with endogenous product entry, direct departure frequencies, and pricing. Impacts are highly asymmetric across carrier types and market segments. Consumer surplus declines by up to 34\% overall and up to 95\% in medium- and long-haul markets. Short-haul markets gain welfare as airlines reallocate capacity toward shorter routes. Airline profits decline by 1--42\%, but carbon tax revenue of \$1.9--2.9 billion and the social value of avoided emissions of \$2.7--4.0 billion more than offset private-surplus losses.
\vspace{0.4cm} \\
\noindent \textbf{Keywords:} Carbon Regulation, Airline Competition, Welfare Analysis
\vspace{0.4cm} \\
\noindent \textbf{JEL Codes:} L93, L13, Q52

\bigskip
\end{abstract}

\thispagestyle{plain}
\end{titlepage}
\setcounter{page}{2}

\section{Introduction}

\noindent The aviation industry's growing share of EU greenhouse gas emissions presents a significant environmental challenge. New-generation aircraft and sustainable aviation fuels (SAFs) are not yet viable at scale, and efficiency gains have historically been outpaced by demand growth. Meeting this challenge motivates market-based policies such as the EU Emissions Trading System (EU ETS). This article investigates how carbon regulation reshapes European airline competition and endogenous route networks. We ask three questions. How do the distinct business models of full-service carriers (FSCs) and low-cost carriers (LCCs) shape their responses to rising carbon costs? How does regulation alter market structure through route entry and exit? And what are the consequences for consumer welfare and its geographic distribution?

The European market differs from its North American and Asia-Pacific counterparts. High population densities and short intercity distances favor direct flights over hub-and-spoke connections and contribute to airport congestion. Europe hosts nearly half of the world's slot-coordinated airports \citep{eurocontrol_atm_us_europe_2024} and features high aircraft utilization. Spanning more than 27 countries and deregulated later than the US market,\footnote{The European Council 
adopted liberalization packages in 1987, 1990, and 1992, creating a substantially liberalized internal market \citep{butcher2010aviation}.} it remains fragmented despite privatization and consolidation over the past 35 years. In 2019, the year of our sample, the market included 14 competing airline groups. Finally, important asymmetries are introduced by the large role of legacy national carriers. Although state aid to national carriers is prohibited except under specified conditions, European FSCs are legacy national or flag carriers, with grandfathered slots, dense hub networks, and legacy costs.

This market structure gives rise to intense competition and a distinction between two dominant business models. FSCs typically operate from congested primary airports, using hub advantages and intercontinental networks to serve both point-to-point and international connecting traffic. LCCs use decentralized point-to-point networks, often from secondary airports, with lower costs and unbundled services. Their wider city portfolios provide more feasible routes for entry. These differences in costs and network organization shape opportunities for profitable route adjustment and so shape the potential impacts of carbon regulation.

Within this competitive asymmetric environment, Europe is implementing stringent aviation carbon policies. Ordinary free allocation of EU ETS allowances to aircraft operators was phased out in 2026, following reductions in 2024 and 2025, although targeted support for eligible SAFs remains \citep{ec_carbon_market_report_2024}. The phase-out coincides with higher jet fuel costs following the 2026 Iran War. Disruptions to Gulf exports and refining capacity have raised crude prices and jet fuel premiums, with industry and energy market assessments suggesting supply pressures could persist into 2027 and beyond. Longer-run pressures include the ReFuelEU mandate to increase SAF usage between 2025 and 2050. SAFs currently cost several times more than conventional fuel and face production shortfalls \citep{easa_eaer_2024}. How these cost shocks interact with airline competition is a first-order question for the industry's future.

We estimate a two-stage game in which airlines first choose direct routes, departure frequencies, and connecting products, then compete on prices. To estimate the game, we use data containing information on European airline networks, prices, and market shares for direct and one-stop services. Our counterfactuals then simulate increasingly stringent carbon taxation through the EU ETS, using an iterative algorithm in which airlines sequentially re-optimize networks in response to higher distance-related costs.

Our estimates reveal substantial cost differences, with LCCs having lower marginal and total costs per passenger.\footnote{We also report results for a smaller regional carrier category comprising Icelandair and Air Europa.} Responses to carbon pricing are highly asymmetric. FSC direct departure frequency declines by 23--62\% and regional frequency by 27--35\%, whereas LCCs primarily reallocate departures toward shorter routes. Medium- and long-haul markets (over 1,500 km) are far more severely affected; short-haul markets gain net welfare as capacity is redirected toward them. Consumers and airlines share the burden. Consumer surplus falls substantially, and profit losses grow with the carbon price. Carbon revenue and the social value of avoided CO\textsubscript{2} emissions more than offset these losses, yielding welfare gains of \$2.2--3.3 billion.

We contribute to two literatures. First, we build on the literature on structural models of airline competition and entry. Most of this literature focuses on the US, where hub-and-spoke networks and connecting traffic are central \citep{berry1990airport,berry1992estimation,berry2010tracing,ciliberto2009market,aguirregabiria2012dynamic,ciliberto2021market,bontemps2023price,yuan2026network}. European studies have examined slot allocation \citep{marra2024market,bauer2025competition} and LCC subsidies \citep{bontemps2024effects}. We build directly on \citet{yuan2026network}'s framework to study how distance-related carbon regulation changes route profitability, reallocates departure capacity, and redistributes welfare. Airport capacity constraints, carriers' feasible redeployment opportunities, and the predominance of direct services distinguish the European setting. Our analysis quantifies how carbon regulation reshapes this imperfectly competitive market through endogenous route entry and exit.

Second, we advance the literature on carbon regulation. Studies of its environmental efficacy include \citet{metcalf2019economics,bayer2020european,colmer2025does,timilsina2022carbon}. We instead examine how carbon pricing reconfigures a large oligopolistic industry. Our approach is closest to \citet{ryan2012costs} and \citet{fowlie2016market} on US cement, and \citet{fowlie2009incomplete} on emissions leakage under imperfect competition. We adapt their insight that environmental policy changes market structure and welfare to aviation. Two features distinguish our setting. Route-specific exposure generates heterogeneous cost shocks, and firms' production locations (i.e. route networks) are endogenous, creating a restructuring channel largely absent in manufacturing.

\textbf{Outline.} Section \ref{sec:background} reviews the industry and data. Section \ref{sec:model} presents the model. Sections \ref{sec:estimation} and \ref{sec:estimation_results} discuss identification, estimation, and results. Section \ref{sec:counterfactual} presents the counterfactuals and Section \ref{sec:conclusion} concludes.

\vspace{-12pt}
\section{Background and Data} \label{sec:background}

\noindent This section provides background on the European airline industry and describes our data sources and processing steps.

\vspace{-12pt}
\subsection{Background: EU Emissions Trading System} \label{sec:background_ets}

\noindent The EU Emissions Trading System (EU ETS), introduced in 2005, establishes a price for the right to emit CO\textsubscript{2} through a cap-and-trade program.\footnote{Earlier market-based environmental policies include US leaded gasoline permits (1982), Icelandic and New Zealand fishing quotas (1984 and 1986), and US sulfur dioxide permits (1995).} It regulates installations across thirty-one countries, covering approximately 40\% of EU greenhouse gas emissions. Entities surrender one European Union Allowance (EUA) per tonne emitted, trade allowances at a uniform European price, and, within limits, bank or borrow across years and phases. The declining cap creates scarcity and an incentive for abatement \citep{ellerman2016european}.

The EU ETS has evolved through several distinct phases. During Phase~I (2005--2007), prices exceeded \euro30 per tonne before collapsing as the cap proved non-binding. Phase~II (2008--2012) coincided with the Great Recession, with prices between \euro8 and \euro30. Phase~III (2013--2020) centralized allocation and shifted toward auctioning while Phase~IV (2021--2030) tightened the cap's annual reduction. Manufacturing industry evidence indicates that the policy reduced emissions 14--16\%, primarily through cleaner technologies, without detectable contractions in economic activity or leakage to unregulated entities \citep{colmer2025does}.

\vspace{-12pt}
\paragraph{Extension to Aviation.} Aviation entered the EU ETS in 2012.\footnote{EU directive 
2008/101/EC included aviation.} Through the end of 2026, the EU ETS and corresponding Swiss and UK systems
collectively covered all flights within the EEA and flights connecting the EEA 
to Switzerland and the UK, subject to specified exemptions. Under the framework, operators must monitor emissions, estimated based on fuel consumption and standard emission factors, and surrender corresponding allowances. Initially, operators were offered free allowances but these were phased out and then completely removed in 2026 \citep{ec_carbon_market_report_2024}.

The EU ETS impact on aviation has several distinctive features. First, the impact is operator and route specific because the carbon charge is proportional to fuel consumption, which depends on aircraft type, passenger load, and distance. Second, the carbon charge affects both fixed per flight operating costs and per passenger marginal costs, with a larger effect on fixed costs since each additional passenger adds little to total weight. Third, carbon prices affect airline entry and exit decisions, and so, unlike manufacturing settings, have first-order effects on network structure and route-level competition.


\vspace{-12pt}
\paragraph{Interaction with Sustainable Aviation Fuel Mandates.} Mandatory SAF adoption adds fuel cost pressures. SAF currently costs two to seven times as much as conventional jet fuel. EASA reports conventional fuel at \euro734 per tonne and biofuel at \euro2,085, projecting a two- to threefold SAF premium through at least 2030 \citep{easa_eaer_2024}. Carbon pricing narrows this gap, but the combined mandates raise operating costs and reinforce the importance of understanding network adjustment.

\vspace{-12pt}
\paragraph{Heterogeneous Impacts across Carrier Types.} 
Carbon regulation does not affect all carriers symmetrically. Because carbon costs scale with distance, all carriers with similar aircraft and loads face similar route-level cost increases. However, network architecture shapes carriers' capacity to respond to these cost increases. LCC point-to-point networks offer flexibility to reallocate departures as relative profitability changes. In contrast, FSC hub-and-spoke networks are anchored at slot-controlled home-country airports. Our model is designed to capture these differences so that we can analyze how higher carbon prices affect entry, exit, network configuration, and pricing.

\vspace{-12pt}
\subsection{Background: European Airline Industry}

\paragraph{The Rise of European Low-Cost Carriers.} Following the deregulation of European aviation in 1992, the European airline industry experienced successive waves of entry, expansion, exit and consolidation. Low-cost carriers (LCCs) expanded through both organic growth and acquisitions. LCCs' share of flights surged from 1.6\% in 1998 to 32.5\% by 2022 \citep{eurocontrol_data_snapshot_2022} and their passenger share reached 56\% by 2016.\footnote{Authors' calculations. Time-series data on passenger shares is not available.} Numerous smaller firms entered but then were absorbed or failed. Some smaller regional carriers operated independently or became subsidiaries of larger carriers.  Full-service carriers (FSCs) responded through cost restructuring, changes to service offerings, and the establishment or acquisition of low-cost subsidiaries. FSC mergers included the formation of Air France–KLM in 2004 and the combination of British Airways and Iberia under IAG in 2011.

The primary strategic difference between FSCs and LCCs lies in their network architecture. FSCs, such as British Airways at London-Heathrow, Iberia at Madrid-Barajas, or Air France at Paris-Charles de Gaulle, typically use a \textit{hub-and-spoke} model to centralize operations, exploiting incumbent hub advantages, grandfathered slot holdings, and economies of scale while funneling passengers from short-haul intra-European flights into lucrative long-haul services to the rest of the world. In contrast, LCCs use a decentralized \textit{point-to-point} (P2P) network focused on high-frequency, short-haul flights between secondary airports. This provides greater routing flexibility by offering direct flights between a wider variety of city pairs.

\begin{figure}[htbp]
    \centering
    \begin{minipage}{0.48\textwidth}
        \centering
        \includegraphics[width=\textwidth]{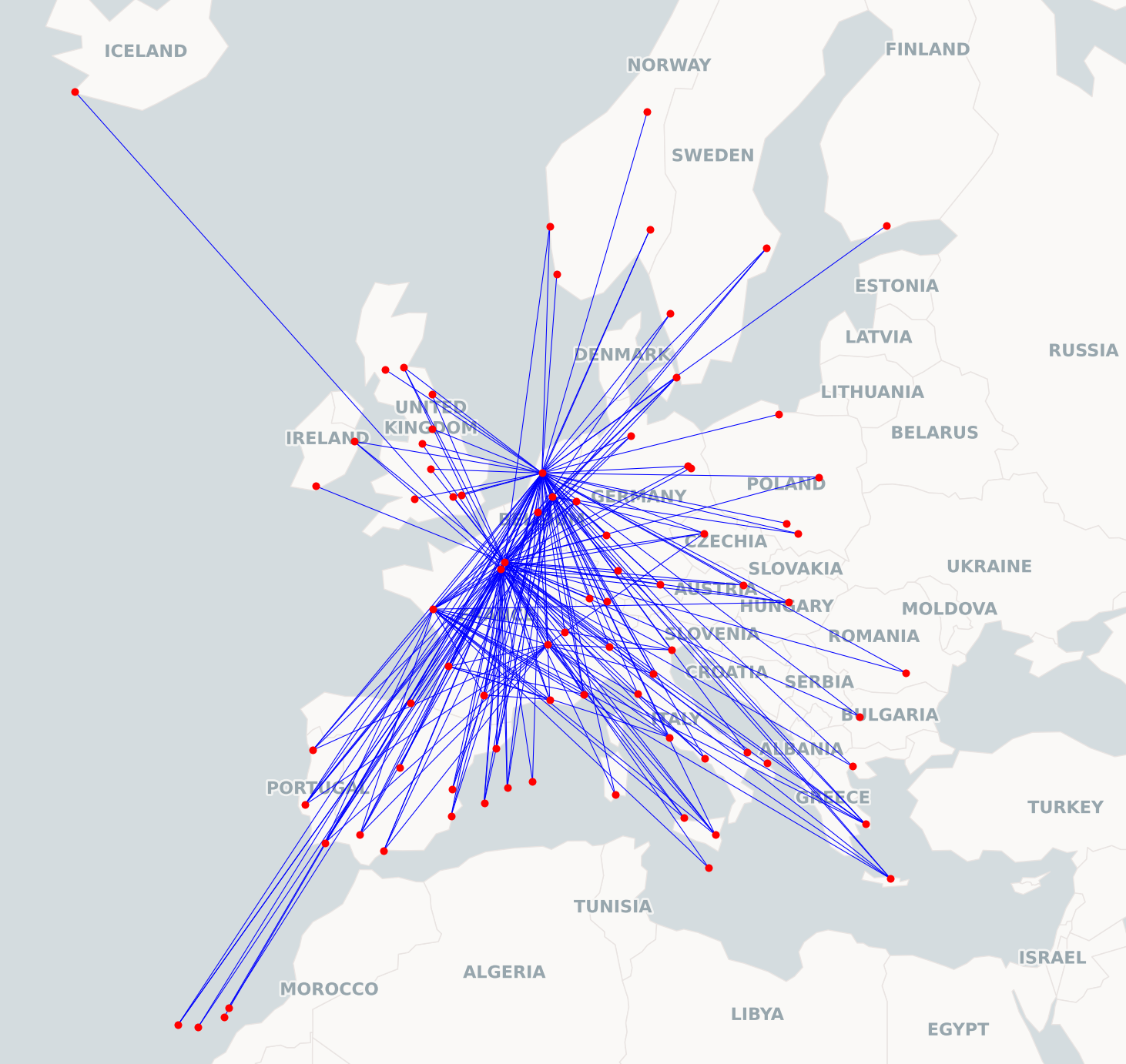}
    \end{minipage}
    \hfill
    \begin{minipage}{0.48\textwidth}
        \centering
        \includegraphics[width=\textwidth]{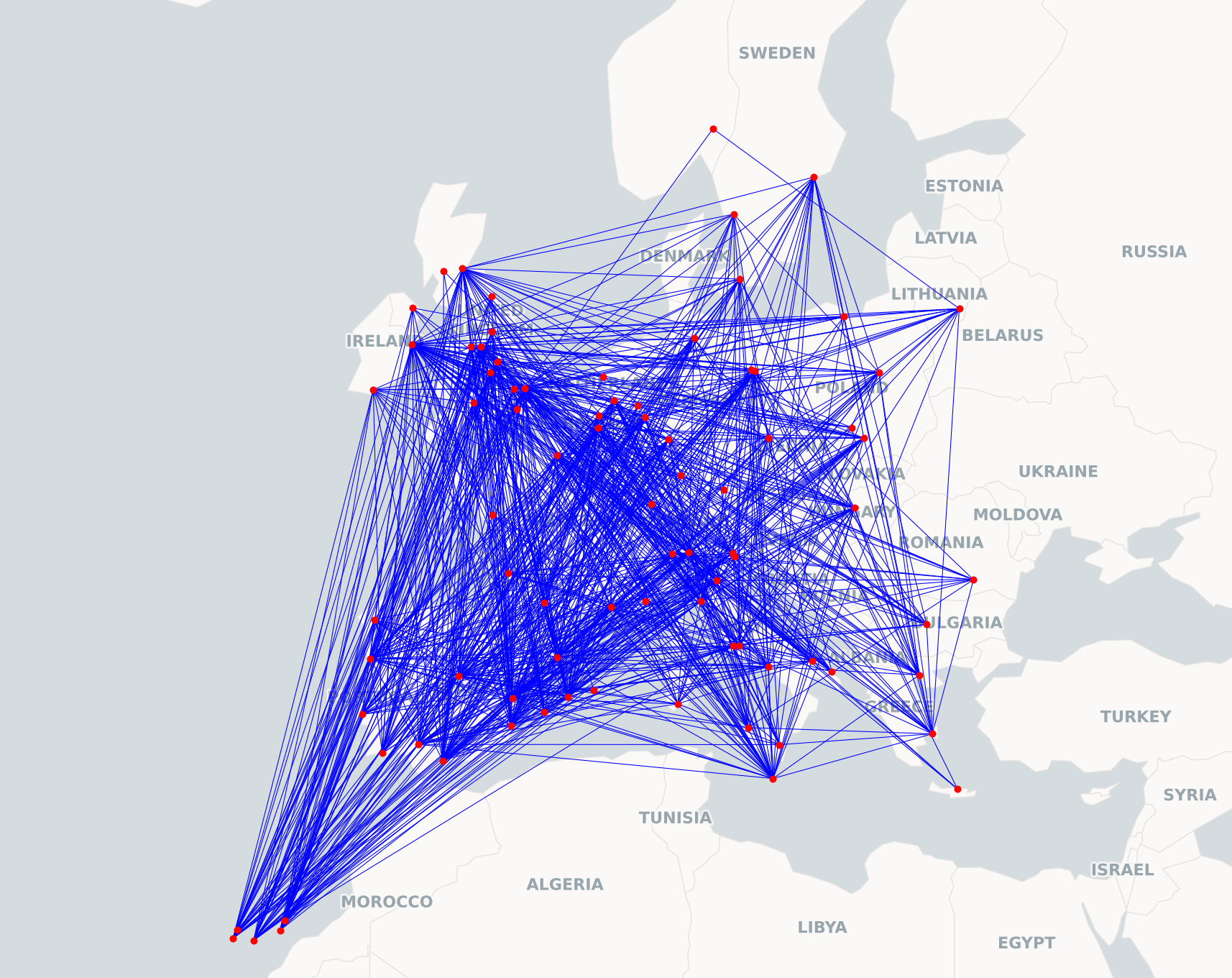}
    \end{minipage}
    \caption{Route Maps of Air France-KLM (left) and Ryanair (right) in Q2 2019}
    \label{fig:figure2}
\end{figure}

The distinction is visually apparent in Figure \ref{fig:figure2}, which contrasts the hub-centric network of Air France-KLM with the diffuse, web-like structure of Ryanair. While Ryanair maintains large operational bases at airports like London Stansted, these do not function as connecting hubs for transfer passengers. Their strategic role is to serve large origin-destination markets, not to facilitate transfers.

In addition, FSC and LCC cost structures differ markedly. FSCs incur higher per-passenger and per-flight costs, driven by higher legacy labor and fleet costs, lower fleet utilization, and premium offerings like business class and meal services. The cost per available seat kilometer for LCCs (excluding fuel) is typically 20\%--30\% lower than for FSCs \citep{doganis2010flying}, granting them a substantial pricing advantage.

Finally, service levels and airport use differ. LCCs unbundle baggage and seat selection, whereas FSCs traditionally offer more inclusive fares.\footnote{Annual reports put LCC ancillary revenue shares at 26--45\%, compared with 11--24\% for FSCs.} FSCs favor major hubs and LCCs smaller secondary airports. In London, Heathrow principally serves FSCs, Gatwick accommodates both types, and Stansted and Luton are major LCC bases. City and Southend serve specialized segments.\footnote{London City mainly serves short-haul business routes; Southend is dominated by charter airlines.} Heathrow's congestion and airport charges make it unattractive to LCCs despite its connectivity. Section~\ref{sec:model} explains how these network, cost, and service differences enter the model.

\vspace{-12pt}
\paragraph{Slot Constraints in European Airports.} 
At congested airports, carriers must acquire slots to gain the right to take off or land at specified times. So, entry or expansion requires acquisition of available slots. By European Council Regulation (EEC) No~95/93, most slots are allocated based on historic precedence. Under the regulation, sufficient use in a given season entitles a carrier to retain its existing slots in the next season. This gives incumbent slot holders, including FSCs with hub operations, an advantage over entrants.\footnote{Historic rights normally require at least 80\% use, subject to exceptions. Our static model neither captures this dynamic incentive nor explicitly allocates airline slot portfolios.} Table~\ref{tab:market_share_slot} documents FSCs' greater presence on routes with slot-controlled endpoints.

\begin{table}[!htbp]
\centering
\caption{Passenger share on routes including at least one slot-controlled airport}
\label{tab:market_share_slot}
\begin{tabular}{lcccc}
\toprule
& \textbf{2016} & \textbf{2017} & \textbf{2018} & \textbf{2019} \\
\midrule
\textbf{Low-cost} & 15.42\% & 14.30\% & 14.54\% & 15.08\% \\
\textbf{Full-service} & 32.91\% & 33.15\% & 32.48\% & 31.96\% \\
\bottomrule
\end{tabular}
\end{table}

To capture the impacts of slot constraints in our estimation and our counterfactuals, we impose capacity constraints at nine airports, requiring capacity to be released before expansion when constraints are binding. Airport and airline-hub controls in demand and marginal costs, and a slot-airport term in fixed costs, also affect route profitability.

\vspace{-12pt}
\paragraph{Airlines and Hubs.} Table~\ref{tab:table1} lists the 14 airline groups, included carriers, and hub or main airports in 2019. We label parents by their largest carrier's name or IATA code. British Airways (BA), for example, represents IAG, including Iberia, Aer Lingus, and Vueling. Its hubs include Heathrow, Madrid-Barajas, and Dublin. Carriers within a parent coordinate through code-sharing and complementary routes. The model distinguishes hubs affecting demand and costs from airport-capacity constraints affecting entry.

\begin{table}[!htbp]
\centering
\caption{Airline groups}
\label{tab:table1}
\begin{threeparttable}
\footnotesize
\setlength{\tabcolsep}{4pt}
\renewcommand{\arraystretch}{1.0}
\begin{tabularx}{\textwidth}{@{}l >{\raggedright\arraybackslash}p{0.20\textwidth} >{\raggedright\arraybackslash}X >{\raggedright\arraybackslash}X@{}}
\toprule
\textbf{Code} & \textbf{Parent group} & \textbf{Included carriers (IATA codes)} & \textbf{Hub/main airports (IATA codes)}\\
\midrule
\multicolumn{4}{@{}l}{\textit{Full-service groups}}\\
BA & International Airlines Group (IAG) & British Airways (BA), BA CityFlyer (CJ), Iberia (IB), Iberia Express (I2), Aer Lingus (EI), Vueling (VY) & London Heathrow (LHR), Madrid Barajas (MAD), Dublin (DUB), Barcelona El Prat (BCN), Rome Fiumicino (FCO)\\
AF & Air France--KLM & Air France (AF), HOP! (A5), KLM (KL), KLM Cityhopper (WA), Transavia (HV/TO), Air Corsica (XK) & Paris Charles de Gaulle (CDG), Amsterdam Schiphol (AMS)\\
LH & Lufthansa Group & Lufthansa (LH), Lufthansa CityLine (CL), Austrian Airlines (OS), Swiss (LX), Brussels Airlines (SN), Eurowings (EW), Air Dolomiti (EN), SunExpress (XQ), SunExpress Deutschland (XG) & Frankfurt (FRA), Munich (MUC), Zurich (ZRH), Vienna (VIE), Brussels (BRU)\\
SK & SAS & Scandinavian Airlines (SK) & Copenhagen (CPH), Stockholm Arlanda (ARN), Oslo Gardermoen (OSL)\\
AY & Finnair & Finnair (AY) & Helsinki (HEL)\\
A3 & Aegean Airlines & Aegean Airlines (A3), Olympic Air (OA) & Athens (ATH)\\
LO & LOT Polish Airlines & LOT Polish Airlines (LO) & Warsaw Chopin (WAW)\\
\midrule
\multicolumn{4}{@{}l}{\textit{Low-cost groups}}\\
DY & Norwegian & Norwegian (DY) & Oslo (OSL) \\
FR & Ryanair Holdings & Ryanair (FR) & London Stansted (STN), Dublin (DUB) \\
LS & Jet2 & Jet2.com (LS) & Leeds Bradford (LBA)\\
U2 & easyJet & easyJet (U2), easyJet Europe (EC) & London Gatwick (LGW)\\
W6 & Wizz Air & Wizz Air (W6) & Budapest (BUD) \\
\midrule
\multicolumn{4}{@{}l}{\textit{Regional groups (study classification)}}\\
FI & Icelandair & Icelandair (FI) & Keflavik (KEF)\\
UX & Air Europa & Air Europa (UX) & Madrid Barajas (MAD)\\
\bottomrule
\end{tabularx}
\begin{tablenotes}[flushleft]\footnotesize
\item \textbf{Note:} The 14 group codes identify the decision-making units in the model. The table lists the 35 carrier codes in the retained 2019 sample using our carrier-to-group mapping. Air Corsica is grouped with AF and SunExpress with LH. HV and TO identify Transavia's Dutch and French operations. Hub/main airports include operating bases and differ from the model's designated connecting airports.
\end{tablenotes}
\end{threeparttable}
\end{table}

\vspace{-12pt}
\paragraph{Airline fleet capacity.} During 2016--2019, European carriers operated near full capacity
with little scope to increase network frequency without expanding 
fleets \citep{eurocontrol_atm_us_europe_2024}. Moreover, fleet expansion was limited. Since 
aircraft orders typically 
require three to five years, in our model we treat fleet capacity as fixed and 
thus require all network and frequency choices to satisfy fleet capacity constraints.

\vspace{-12pt}
\subsection{Data}

\noindent Our data come from Sabre Market Intelligence \citep{sabre_market_intelligence}, a reservation and pricing system used by major European airlines for fare optimization. The data cover 2016--2022. We focus on 2019, the latest pre-COVID year.

Quarterly airline-itinerary observations record endpoint and connecting airports, non-stop status, average fares, frequency, travel time, and passengers. We retain direct and one-stop services in the top-100-airport sample, spanning 83 cities. Following \citet{yuan2026network} and \citet{bontemps2023price}, we aggregate opposite directions, which typically have similar prices and frequencies. A product is defined to be an itinerary offered by a parent group. A market is the set of all products joining a pair of endpoint cities. Distance is total itinerary distance, including both connecting legs. Direct-flight itineraries carry 99.0\% of retained passengers, but one-stop flights matter for substitution and the profitability of component legs. Eurostat metropolitan populations determine market size \citep{eurostat_cities_database}.

Table \ref{tab:table2} presents summary statistics. Our estimation sample contains the 14 parent airline groups: seven full-service, five low-cost, and two regional groups (Table \ref{tab:table1}). The six largest groups are the three primary FSCs (British Airways, Air France, and Lufthansa) and the three primary LCCs (Ryanair, easyJet, and Wizz Air). These six groups account for 87.2\% of passenger traffic in the sample. We observe 13,509 product-quarters across 2,141 city pairs and 7,647 city-pair-quarters. Of these, 11,487 product-quarters are direct flights and 2,022 are one-stop flights. The average fare is \$91.44, the average direct-flight frequency is 2.88 directional departures per day, and the average itinerary distance is about 1,482 kilometers. In total, the airlines in our 2019 sample served 348.26 million passengers.

\begin{table}[!htbp]
\centering
\begin{threeparttable}
\caption{Summary statistics}
\label{tab:table2}
\begin{tabular}{@{}l r @{\hspace{1.5em}} l r r@{}}
\toprule
\multicolumn{2}{l}{\textbf{(a) Market dimensions}} & \textbf{(c) Demand and cost} & \textbf{Mean} & \textbf{St.Dev}\\
\midrule
Product-quarters & 13,509 & Fare (100 USD) & 0.91 & 0.58\\
City pairs & 2,141 & Frequency (daily) & 2.88 & 4.03\\
Passengers (million) & 348.26 & Distance (1,000 km) & 1.48 & 0.73\\
 & & Market size (1 million) & 2.36 & 1.66\\
\bottomrule
\end{tabular}
\par\smallskip
\begin{tabular}{@{}lccccccc@{}}
\toprule
\multicolumn{8}{@{}l}{\textbf{(b) Market shares}}\\
 & \multicolumn{3}{c}{FSCs} & \multicolumn{3}{c}{LCCs} & Other\\
\cmidrule(lr){2-4}\cmidrule(lr){5-7}
 & \makecell{British\\Airways} & Air France & Lufthansa & Ryanair & EasyJet & Wizz Air & \\
Share & 0.15 & 0.09 & 0.13 & 0.25 & 0.21 & 0.05 & 0.13\\
\bottomrule
\end{tabular}
\end{threeparttable}
\end{table}

Table \ref{tab:table3} reports city-level connectivity and frequency statistics for the six largest airline groups, the three largest FSCs and the three largest LCCs. Panels (a) and (b) reveal that FSCs generally maintain greater connectivity and higher frequencies at their hubs. For instance, Lufthansa Group operates an average of 7.8 daily departures per direct route serving Frankfurt, compared with 3.1 for Ryanair routes serving Dublin. This contrast is also evident in Panels (c) and (d), which measure network concentration. Approximately 46\% of Air France-KLM's direct-flight frequency is to or from Amsterdam, with a similar share to or from Paris, a clear empirical signature of a hub-and-spoke model. LCCs generally exhibit lower concentration levels, with their routes more evenly distributed across a wide range of cities, reflecting their decentralized point-to-point strategy.

\begin{table}[!htbp]
\centering
\caption{Connectivity and frequency: six largest airline groups, two most connected cities}
\label{tab:table3}
{\footnotesize
\begin{tabular}{c c c c c c c}
\toprule
Group & First City & Cities Served & Freq. & Second City & Cities Served & Freq.\\
\midrule
\textbf{(a) Full service:} & & & & & &\\
BA & London & 67 & 7.4 & Barcelona & 60 & 4.8\\
AF & Amsterdam & 71 & 7.0 & Paris & 63 & 6.5\\
LH & Frankfurt & 61 & 7.8 & Munich & 60 & 7.4\\
\midrule
\textbf{(b) Low Cost:} & & & & & &\\
FR & Dublin & 59 & 3.1 & London & 56 & 4.7\\
U2 & London & 61 & 7.2 & Berlin & 56 & 3.9\\
W6 & Budapest & 40 & 1.5 & Bucharest & 27 & 1.5\\
\midrule
Group & First City & Concentration & & Second City & Concentration &\\
\midrule
\textbf{(c) Full service:} & & & & & &\\
BA & London & 46.2\% & & Barcelona & 21.0\% &\\
AF & Amsterdam & 46.3\% & & Paris & 46.5\% &\\
LH & Frankfurt & 28.5\% & & Munich & 25.8\% &\\
\midrule
\textbf{(d) Low Cost:} & & & & & &\\
FR & Dublin & 12.6\% & & London & 18.9\% &\\
U2 & London & 32.1\% & & Berlin & 15.5\% &\\
W6 & Budapest & 22.0\% & & Bucharest & 17.3\% &\\
\bottomrule
\end{tabular}
}
\par\smallskip
\begin{minipage}{\linewidth}\footnotesize
\textbf{Note:} The table presents statistics for the six largest airline groups and their two most connected cities. Cities are ranked by the number of directly connected cities. Cities Served counts directly connected cities. Freq. is the average number of daily departures per direct route serving the listed city. Concentration measures the proportion of direct flight frequencies to/from the hub city relative to all the airline's direct flight frequencies.
\end{minipage}
\end{table}

Table \ref{tab:combined_competition}(a) shows that around 47.5\% of market-quarters in 2019 are served by more than one airline group. Average fares are slightly lower in two- and three-firm markets than in monopoly markets, although the relationship is not monotone. Fares are averaged within market-quarters, and the standard deviation of these averages is highest in monopoly markets. Table \ref{tab:combined_competition}(b) shows that, in 2019, FSCs have about 1.6 times as many slot-airport route-quarter observations per parent airline as LCCs. Table \ref{tab:combined_competition}(c) reports average quarterly changes in route observations per parent airline over 2016--2019. Both FSCs and LCCs expand in Q2, with increases of 26.3 and 44.0 route observations, respectively.

\begin{table}[!htbp]
\centering
\caption{Competition and Network Statistics}
\label{tab:combined_competition}
\footnotesize
\begin{subtable}{\textwidth}
\centering
\caption{Price and competition}
\begin{tabular}{lcccccc}
\toprule
\textbf{Number of Firms} & \textbf{1} & \textbf{2} & \textbf{3} & \textbf{4} & \textbf{5} & \textbf{6}\\
\midrule
\textbf{Market-quarters} & 4,015 & 2,567 & 809 & 224 & 30 & 2\\
\textbf{Percentage} & 52.50\% & 33.57\% & 10.58\% & 2.93\% & 0.39\% & 0.03\%\\
\textbf{Average Fare (\$100)} & 0.901 & 0.897 & 0.882 & 0.995 & 1.034 & 0.787\\
\textbf{Std. Dev.} & 0.626 & 0.416 & 0.298 & 0.283 & 0.320 & 0.009\\
\bottomrule
\end{tabular}
\end{subtable}

\vspace{0.2cm}
\begin{subtable}{\textwidth}
\centering
\caption{Slot-Airport Exposure}
\begin{threeparttable}
\begin{tabular}{lcccc}
\toprule
& \textbf{2016} & \textbf{2017} & \textbf{2018} & \textbf{2019}\\
\midrule
\textbf{Low-cost} & 201 & 214 & 233 & 239\\
\textbf{Full-service} & 383 & 384 & 392 & 389\\
\bottomrule
\end{tabular}
\footnotesize
\textbf{Note:} Annual number of route-quarter observations with at least one slot-controlled airport per parent airline.
\end{threeparttable}
\end{subtable}

\vspace{0.2cm}
\begin{subtable}{\textwidth}
\centering
\caption{Quarterly dynamics}
\begin{threeparttable}
\begin{tabular}{llcccc}
\toprule
& & \textbf{Q1} & \textbf{Q2} & \textbf{Q3} & \textbf{Q4}\\
\midrule
\multirow{2}{*}{\textbf{All Routes}}
& Low-cost & $-13.9$ & 44.0 & 10.8 & $-27.9$\\
& Full-service & $-9.9$ & 26.3 & 8.3 & $-21.7$\\
\midrule
\multirow{2}{*}{\textbf{Slot-Airport Routes}}
& Low-cost & $-1.3$ & 5.4 & 2.0 & $-4.0$\\
& Full-service & $-5.0$ & 12.0 & 3.4 & $-9.4$\\
\bottomrule
\end{tabular}
\footnotesize
\textbf{Note:} Average quarterly change in route observations per parent airline over 2016--2019.
\end{threeparttable}
\end{subtable}
\end{table}

\begin{figure}[htbp]
    \centering
    \includegraphics[width=1\textwidth]{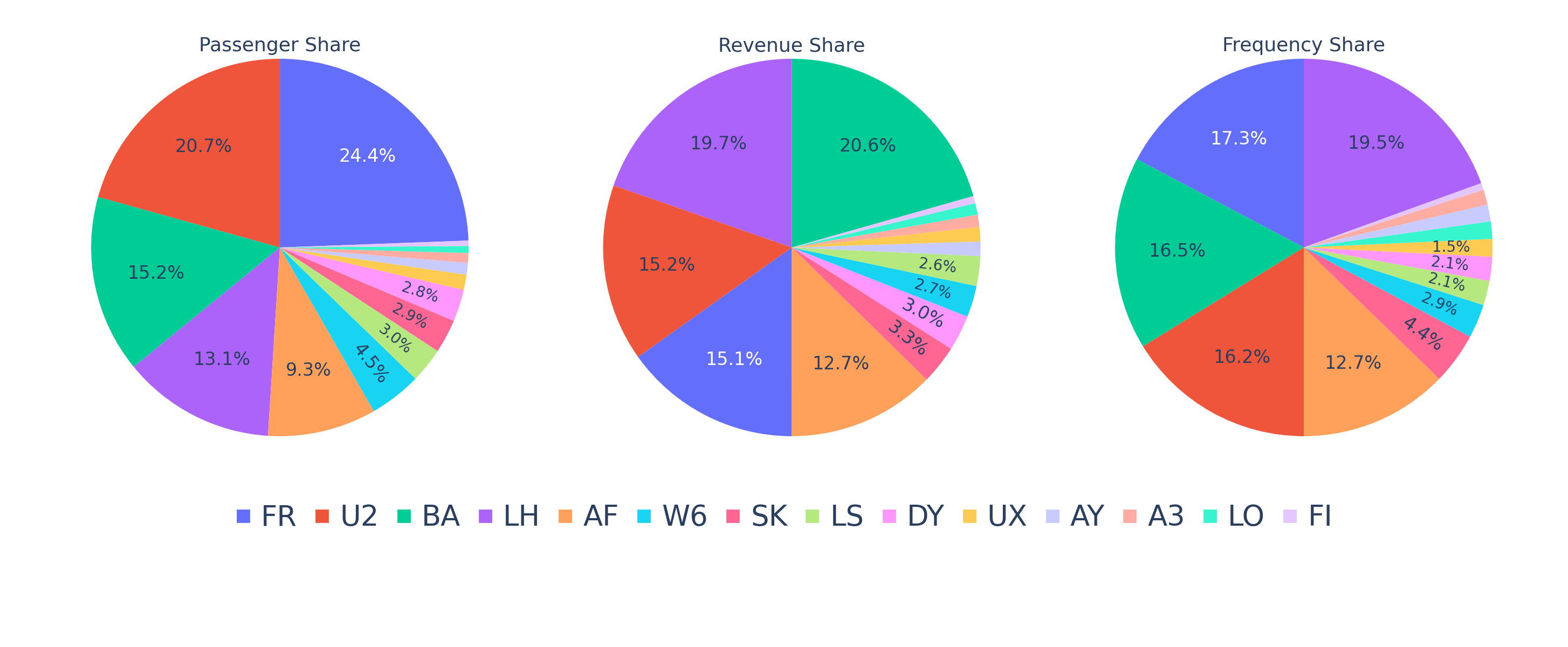}
    \caption{Passenger, revenue, and frequency shares for all 14 parent airline groups, Q2 2019. Group codes are defined in Table \ref{tab:table1}}
    \label{fig:figure8}
\end{figure}

Figure \ref{fig:figure8} shows passenger, revenue, and frequency shares for all 14 parent airline groups. Their codes are defined in Table \ref{tab:table1}. While Ryanair (FR) and easyJet (U2) have the largest passenger shares, IAG (BA) and Lufthansa Group (LH) have the highest revenue shares. Lufthansa also has the highest frequency share, followed by Ryanair and IAG, reflecting differences in fares and schedule intensity.

\begin{figure}[htbp]
    \centering
    \begin{minipage}{0.8\textwidth}
    \begin{subfigure}{0.49\textwidth}
        \includegraphics[width=\linewidth,trim=0bp 61.2bp 972bp 0bp,clip]{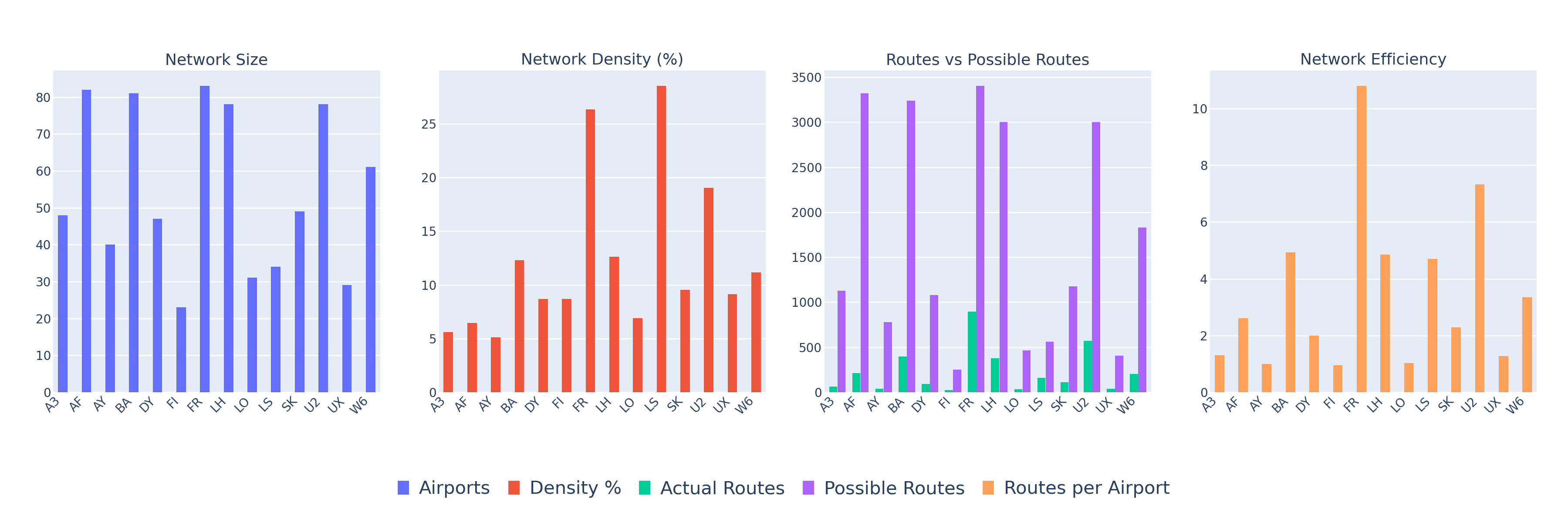}
        \caption{Network size (airports)}
    \end{subfigure}\hfill
    \begin{subfigure}{0.49\textwidth}
        \includegraphics[width=\linewidth,trim=324bp 61.2bp 648bp 0bp,clip]{figures/network_density.png}
        \caption{Network density (\%)}
    \end{subfigure}
    \par\medskip
    \begin{subfigure}{0.49\textwidth}
        \includegraphics[width=\linewidth,trim=648bp 61.2bp 324bp 0bp,clip]{figures/network_density.png}
        \caption{Actual (green) and possible (purple) routes}
    \end{subfigure}\hfill
    \begin{subfigure}{0.49\textwidth}
        \includegraphics[width=\linewidth,trim=972bp 61.2bp 0bp 0bp,clip]{figures/network_density.png}
        \caption{Network efficiency (routes per airport)}
    \end{subfigure}
    \end{minipage}
    \caption{Network size is the total number of airports served. Network density is the number of airport pairs served by direct flights divided by the total number of possible pairs. Panel (c) displays actual pairs linked by direct flights and potential pairs. Network  efficiency is the number of direct routes per airport. Group codes are defined in Table \ref{tab:table1}}
    \label{fig:figure9}
\end{figure}

Figure~\ref{fig:figure9} reports statistics on network size, density, and efficiency. Network size is defined to be the number of airports served. In our model, this determines the scope of feasible entry alternatives. BA, AF, LH, FR, U2, and W6 have the largest networks. Network density is the number of active direct airport pairs divided by possible pairs. FSCs have lower density, consistent with hub-and-spoke networks. Network efficiency is a measure of the average number of active direct routes per airport. Ryanair serves over ten on average compared with fewer than three for Air France.

\vspace{-12pt}
\section{Model} \label{sec:model}

\noindent This section introduces a static two-stage model of airlines' entry, direct departure frequency, and pricing decisions. In the first stage, airlines simultaneously choose direct routes and their departure frequencies, together with the connecting flights to offer. In the second stage, airlines compete on prices to attract customers. Our model closely follows the framework of \citet{yuan2026network}, including the linkage between direct and connecting services and the use of revealed-preference inequalities to estimate costs. We modify their framework for the European context by retaining multiple airports per city and multiple options for connecting airports and by incorporating capacity-constrained redeployment when analyzing airlines’ responses to distance-related carbon charges.

Let $g \in \mG$ be an airline group and let a market $m \in \mM$ be defined by a non-directional city-pair $c,d \in \mC$ where $\mC$ is the set of cities.\footnote{The definition of a market as a city-pair follows \citet{berry1992estimation,aguirregabiria2012dynamic,yuan2026network}.} An itinerary, also referred to as a route, is defined by its endpoint airports, an indicator for non-stop or one-stop service, and, for one-stop service, its connecting airport. We aggregate opposite travel directions, so endpoint airport pairs are non-directional. A direct or non-stop route $r\in\mR$ is an airport pair $a,b\in\mA$, where $\mA$ is the set of airports. A connecting or one-stop route additionally specifies the connecting airport $h$. A product $j$ is a route or itinerary offered by a particular airline group. We use ``direct'' and ``non-stop'' interchangeably, and ``connecting'' (or ``indirect'') to denote a one-stop itinerary comprising two direct legs. An airline group's direct operating network consists of its active direct routes and their departure frequencies. Indirect flights use this network and require both component legs to be active, but offering an indirect service is a separate decision. In general prose, a flight may refer to a direct or connecting service. When counting aircraft operations, we use ``departures'' to distinguish them from itineraries or products. Furthermore, let $j=0$ denote the outside option of not flying. Let $\mJ_{gm}$ be the set of products chosen by airline $g$ in market $m$ in stage one of the two-stage game, and let the vector $\mathbf{N}_g$ represent airline $g$'s product-offering vector, with $N_{gj}=1$ if product $j$ is offered. In the pricing stage, the set of products available in market $m$ is the outside option $j=0$ plus $\mJ_m=\bigcup_g\mJ_{gm}$. That is, it is the outside option plus the set of products chosen by airlines in stage one. Denote the number of products in market $m$ by $J_m=|\mJ_m|$. Note that an airline may offer multiple products in a market by using different airports in the same city or by offering connecting itineraries that use different connecting airports.

\vspace{-12pt}
\paragraph{Airlines' first-stage choices and consideration sets.} Each airline's first-stage choices include choice of product portfolio $\mathbf{N}_{g}$ and direct departure frequencies $\mathbf{F}_{g}$. For the product portfolio choice, we impose the following constraints on firms' consideration sets. We assume that, in the short run, an airline can only enter routes between cities it already serves, as in \cite{berry1992estimation}. Costs to expand to unserved cities are assumed to be too high to be profitable in the short run. The city set is fixed at its observed quarterly support throughout the simulation. At the nine airports treated as capacity constrained, aggregate direct departure frequency cannot exceed observed capacity.\footnote{These airports are Amsterdam, Athens, Paris Charles de Gaulle, Copenhagen, Frankfurt, Helsinki, London Heathrow, Madrid, and Munich.} Thus entry into an airport with fully occupied capacity requires departure capacity to be released elsewhere at that airport. Finally, to capture legacy entry costs of flag carriers, we require that, when considering redeploying departures from an observed direct route to an alternative route, the number of alternative route endpoints located in the airline's home country is weakly greater than the number on the observed route. These consideration set constraints capture the assumption that expanding services in directions outside the support of the observed route network entails greater costs. We assume these greater costs are sufficiently large that such entry is not feasible in our estimation nor in our counterfactual simulation. We use $\mathcal{J}_{g}^{feas}$ to denote the resulting consideration set of feasible products for airline $g$, including direct flights and one-stop flights supported by feasible direct legs. Regarding the choice of direct departure frequency $\mathbf{F}_{g}$, total direct frequency across an airline's network is bounded above by its observed quarterly total, with released departure capacity available for redeployment.

The model represents the three distinctions between FSCs and LCCs discussed in Section 2.2 through several channels. First, differences in network architecture enter through airlines’ observed city portfolios, designated connecting airports, and the dependence of connecting services on their component direct legs. These features shape feasible redeployment opportunities and the network-wide profit consequences of changing a route. Flag carriers additionally face the home-country restriction described above. Second, differences in costs are captured by airline-quarter effects and product characteristics in marginal costs, while fixed operating costs vary with direct departure frequency, distance, market size, and capacity-constrained airport exposure. The fixed-cost coefficients are common across carriers, so differences in their implied fixed costs arise from their networks and operating choices rather than separate FSC and LCC cost parameters. Third, differences in service offerings and airport use enter through airline-quarter, major-airport, and airline-hub effects and other product controls in demand and marginal costs, together with airport-capacity constraints on entry. These controls capture systematic differences across carriers and products, although we do not separately model baggage charges, individual service amenities, or a secondary airport cost advantage.

\vspace{-12pt}
\subsection{Second Stage: Demand and Pricing}

\noindent In the second stage, given product portfolios and direct departure frequencies, airlines compete in prices. They simultaneously set prices for all products in each market to maximize profits under complete information.

\vspace{-12pt}
\paragraph{Demand.} The demand model is a discrete choice model following \cite{berry2010tracing} and \cite{yuan2026network}. For a product $j$ in market $m$, the utility of consumer $i$ is given by:
\begin{equation*}
U_{ijm}=
\begin{cases} 
    -\alpha p_{jm} + x_{jm}\beta + \xi_{jm} + \nu_{im}(\lambda) + \lambda\varepsilon_{ijm} & \text{if } j \in \mJ_{m} \\
    \varepsilon_{i0m} & \text{if } j=0
   \end{cases}
\end{equation*}
where $x_{jm}$ is a vector of product characteristics, $p_{jm}$ is the product price, $\xi_{jm}$ is the unobserved (to researchers) product characteristic, $\nu_{im}(\lambda)$ is the ``nested logit'' shock, $\varepsilon_{ijm}$ is the i.i.d.\ extreme value type I utility shock, $\alpha$ is the price coefficient, $\beta$ is the vector of utility parameters,  and $\lambda \in (0,1)$ is the nesting parameter. Let $\theta_{d} = (\alpha,\beta, \lambda)$ denote the vector of demand parameters.

The product characteristics $x_{jm}$ include a constant, the logarithm of product frequency, total itinerary distance and its square, a non-stop indicator, a high-speed rail competition indicator, airline-quarter fixed effects, major airport fixed effects, airline-hub fixed effects, and fixed effects for the top 50 cities ranked by population. Total itinerary distance equals direct distance for non-stop flights and the sum of the two leg distances for one-stop flights. It can therefore vary across itineraries in the same market. We also include indicators for routes with two small-market endpoints or two large-market endpoints, and their interactions with Aegean, LOT, Air France-KLM, and Finnair. Mixed endpoint categories are the reference group. Frequency and distance coefficients are common across airlines. Higher frequency offers consumers more travel options and greater flexibility. The airline-hub fixed effect is important because major European hubs serve both intra-European passengers and a substantial volume of extra-European transfer passengers. Higher levels of demand due to these additional extra-European passengers are captured by these airline-hub fixed effects.\footnote{International transfer passengers traveling on single itineraries with short layovers are not included in our dataset. However, the intra-European legs of international passenger journeys with long layovers are included.}

The model implies that the market share for product $j$ in market $m$ is:
\begin{align*}
 s_{jm}(\mathbf{p}_{m}, \mathbf{x}_{m}, \bm{\xi}_{m}; \theta_d) =&
    \underbrace{\frac{(\sum_{k \in \mJ_{m}}\exp((-\alpha p_{km} + x_{km} \beta + \xi_{km})/\lambda))^{\lambda}}{1 + (\sum_{k \in \mJ_{m}}\exp(( - \alpha p_{km} + x_{km} \beta + \xi_{km})/\lambda))^{\lambda}}}_{\text{Probability of flying}}
    \\
&    \times
    \underbrace{\frac{\exp(( - \alpha p_{jm} + x_{jm} \beta + \xi_{jm})/\lambda)}{\sum_{k \in \mJ_{m}}\exp(( - \alpha p_{km} + x_{km} \beta + \xi_{km})/\lambda)}}_{\text{Conditional probability of choosing } j}
\end{align*}
where $\mathbf{p}_{m} = (p_{km}: k \in \mJ_{m})$, $\mathbf{x}_{m} = (x_{km}: k \in \mJ_{m})$, and $\bm{\xi}_{m} = (\xi_{km}: k \in \mJ_{m})$. Taking the logarithm, the demand equation can be expressed as the following linear equation:
\begin{equation} \label{eq:demand}
    \ln(s_{jm}) - \ln(s_{0m}) = -\alpha p_{jm} + x_{jm} \beta + (1-\lambda) \ln(s^{*}_{jm}) + \xi_{jm}
\end{equation}
where $s_{0m}$ is the market share of the outside option, and $s^{*}_{jm}$ is the within-nest market share of product $j$. The endogenous variables in this equation are the prices $p_{jm}$ and the within-nest market shares $s^{*}_{jm}$.

\vspace{-12pt}
\paragraph{Supply.} Airlines simultaneously set prices in each market to maximize profits:
\begin{equation*}
    \sum_{m \in \mM} \sum_{j \in \mJ_{gm}} (p_{jm} - \text{MC}_{jm}) \cdot s_{jm}(\mathbf{p}_{m}, \mathbf{x}_{m}, \bm{\xi}_{m}; \theta_d) \cdot \text{MS}_{m} \hspace{0.5cm} \forall g
\end{equation*}
where $\text{MC}_{jm}$ is the marginal cost of product $j$ in market $m$ and $\text{MS}_{m}$ is the market size defined as the geometric mean of the populations of the two endpoint cities. Let $\mathbf{O}_{m}$ be the ownership matrix for market $m$ where element $(j,k)$ equals 1 if the same firm owns both products $j$ and $k$. The Bertrand--Nash F.O.C.s for profit maximization yield:
\begin{equation*}
    \text{MC}_{m} = \mathbf{p}_{m} + (\mathbf{O}_m \odot \frac{\partial s_m}{\partial p_m})^{-1} s_m
\end{equation*}
where $\text{MC}_{m}$ is a $J_m \times 1$ vector of marginal costs for all products in market $m$, and $\odot$ denotes the element-wise product.

We assume marginal cost is a function of observable product characteristics:
\begin{equation}
    \text{MC}_{jm} = x_{jm} \theta_{s} + \omega_{jm} \label{eq:marginal_cost}
\end{equation}
where $\omega_{jm}$ is an unobserved cost shock and $\theta_{s}$ is a marginal cost parameter vector. The marginal cost regression uses the same exogenous characteristics and fixed effects as demand, including the non-stop indicator and total itinerary distance. When we compute our counterfactual simulations, we assume increased fuel costs increase marginal cost per passenger per kilometer as detailed in Section \ref{sec:counterfactual} below.

\vspace{-12pt}
\subsection{First Stage: Entry and Frequency}

\noindent In the first stage, airlines simultaneously choose product portfolios and direct departure frequencies. Airlines incur fixed operating costs for each active itinerary. These include operating costs for direct itineraries and incremental offering costs for connecting itineraries. For airline $g$, offering product portfolio $\mathbf{N}_g$ and direct departure frequency $\mathbf{F}_g$, we assume that the total fixed operating cost is:
\begin{equation*}
    \text{FC}_g(\mathbf{N}_{g}, \mathbf{F}_{g}, \bm{\kappa}_{g} ; \theta_{fc}) = \sum_{j \in \mathcal{J}_g^{feas} } 
    N_{gj} \cdot \left( z_j(f_{j}) \theta_{fc} + \kappa_{j}(f_{j}) \right)
\end{equation*}
where $z_j(f_j)$ is a vector of observable product characteristics including a constant, frequency times distance, the logarithm of market size, the number of capacity-constrained airports on the route, and an indicator for one-stop itineraries. To be specific, the five-dimensional vector $z_j$ equals $(1,f_j d_j,\log\text{MS}_m,S_j,0)$ for a direct flight and $(0,0,0,0,1)$ for an indirect flight, where $d_j$ is direct distance in thousands of kilometers and $S_j$ counts capacity-constrained endpoint airports. The specification captures the fact that adding an indirect service incurs only an incremental cost since the direct leg costs are already counted. The variable $\kappa_j(f_j)$ is an unobserved fixed cost shock, and $\theta_{fc}$ is a vector of fixed cost parameters. For direct flights the shock is specific to airline, quarter, route, and frequency. For indirect flights it is specific to airline, quarter, endpoints, and connector, and does not vary with frequency. $\bm{\kappa}_g$ is the vector of these shocks for airline $g$. Note that fuel costs, which are impacted by carbon prices, are a fixed cost of operating a direct route and are proportional to frequency times distance. The frequency-distance coefficient includes these fuel costs and other distance-related fixed operating costs. When we compute our counterfactual simulations, increases in EU ETS prices add to this component without imposing the same aircraft charge a second time on connecting products. This is detailed in Section \ref{sec:counterfactual} below.

We assume that firms choose their product portfolios in stage one before the second-stage shocks $\xi_{jm}$ and $\omega_{jm}$ are realized.\footnote{Prior work, including \cite{aguirregabiria2012dynamic}, \cite{sweeting2013dynamic}, \cite{eizenberg2014upstream}, and \cite{yuan2026network}, makes an analogous assumption.} Let $\left(\mathbf{N},\mathbf{F},\mathbf{X} \right)$ be the networks, frequencies, and product characteristics of all airlines in all markets. Then, for each airline $g$, expected second-stage profits can be written as:
\begin{equation*}
    \Pi_{2g}(\mathbf{N}, \mathbf{F}, \mathbf{X}; \theta_d, \theta_{s}) = \bE_{\bm{\xi}, \bm{\omega}}\left[ \sum_{m \in \mM} \sum_{j \in \mJ_{gm}} (p_{jm} - \text{MC}_{jm}) \cdot s_{jm}(\mathbf{p}_{m}, \mathbf{x}_{m},\bm{\xi}_m; \theta_d) \cdot \text{MS}_{m}\right]
\end{equation*}
In this expression, $\mathbf{p}_{m}$ is the equilibrium price vector that arises in stage two in market $m$ after the demand and cost shocks $\left(\bm{\xi}_{m},\bm{\omega}_m \right)$ are realized. The expectation is taken over all unobserved demand and cost shocks in all markets. We assume airlines know the distributions of these shocks when making entry and frequency decisions.

Airlines observe all fixed cost shocks and simultaneously choose $(\mathbf{N}_{g}, \mathbf{F}_{g})$ to maximize expected net profit, $\Pi_{2g}-\text{FC}_{g}$.

\vspace{-12pt}
\subsection{Equilibrium}

\noindent Airlines choose product portfolios and direct departure frequencies in stage 1 and prices in stage 2. In this two-stage game, we consider pure strategy subgame perfect equilibria in product portfolios, direct departure frequencies, and prices: $\{\mathbf{N}^{*}, \mathbf{F}^{*}, \mathbf{P}^{*}\}$. The existence and uniqueness of equilibrium in the second-stage pricing game are established by \cite{nocke2018multiproduct} for multi-product nested logit models. However, equilibrium in the first-stage game is not guaranteed to exist, as noted by \cite{bontemps2023price} and \cite{yuan2026network}. We assume the existence of a first-stage equilibrium, but do not assume uniqueness and allow for multiple equilibria.

\vspace{-12pt}
\section{Identification and Estimation Strategies} \label{sec:estimation}

\noindent There are two endogenous variables in \eqref{eq:demand}, price and within-nest share. We use three excluded instruments: the average frequency of other products in the market; the number of distinct airline parent groups in the market; and an indicator that at least one endpoint airport is a hub for the airline's alliance.

We assume that, conditional on the included controls, these instruments are orthogonal to the second-stage demand shocks. The justification for this assumption is as follows. Network structure and direct-flight frequency are chosen several months before prices. When airlines make these choices, they forecast profits based on persistent demand and cost shifters including city, airport, and hub fixed effects. Their forecast errors are driven by high-frequency, hard-to-predict demand and cost shocks. We assume that these demand and cost shocks are orthogonal to the variables in their information sets. This restriction allows network and frequency choices to respond to predictable demand and cost shifters while excluding correlation with the unobservable innovations that later affect costs and demand.

Based on this identification strategy, demand is estimated by instrumental variable GMM. Marginal costs are recovered from the pricing first-order conditions, after which we estimate equation \eqref{eq:marginal_cost} by least squares using the same exogenous controls as in demand.

Identification of the fixed cost parameters $\theta_{fc}$ requires more discussion. These parameters are set identified based on moment inequalities arising from the firms' entry/exit decisions.  

\vspace{-12pt}
\paragraph{Construction of Moment Inequalities.} To ease notation, we suppress dependence on $(\mathbf{X},\theta_s,\theta_d)$ and on the competitors' strategies. Let $\Pi_{1g}(\mathbf{N}_g,\mathbf{F}_g, \bm{\kappa}_g ; \theta_{fc}) := \Pi_{2g}(\mathbf{N}_g, \mathbf{F}_g) - \text{FC}_{g}(\mathbf{N}_{g}, \mathbf{F}_{g}, \bm{\kappa}_{g} ; \theta_{fc})$ denote airline $g$'s stage 1 profit conditional on its own actions, its competitors' actions, and all other state variables. Assuming observed choices $(\mathbf{N}_g^*,\mathbf{F}_g^*)$ maximize profits, alternative feasible actions $(\mathbf{N}^{a}_{g},\mathbf{F}^{a}_{g})$ must not increase profits. That is:
\begin{equation*}
    \Pi_{1g}(\mathbf{N}^{*}_{g}, \mathbf{F}^{*}_{g}, \bm{\kappa}_{g} ; \theta_{fc}) - \Pi_{1g}(\mathbf{N}^{a}_{g}, \mathbf{F}^{a}_{g}, \bm{\kappa}_{g} ; \theta_{fc}) = \Delta \Pi_{1g} (\mathbf{N}^{*}_{g}, \mathbf{F}^{*}_{g}, \mathbf{N}^{a}_{g}, \mathbf{F}^{a}_{g} ; \theta_{fc}) - \Delta\kappa_g^a \geq 0
\end{equation*}
where $\Delta \Pi_{1g} (\mathbf{N}^{*}_{g}, \mathbf{F}^{*}_{g}, \mathbf{N}^{a}_{g}, \mathbf{F}^{a}_{g} ; \theta_{fc})$ is the deterministic component of the difference in profit and $\Delta\kappa_g^a$ is the sum of fixed cost shocks under the observed network minus the corresponding sum under the alternative network. Under the linear fixed cost specification, the deterministic component can be written:
\begin{equation*}
    \Delta \Pi_{1g} (\mathbf{N}^{*}_{g}, \mathbf{F}^{*}_{g}, \mathbf{N}^{a}_{g}, \mathbf{F}^{a}_{g} ; \theta_{fc}) = \Pi_{2g}(\mathbf{N}_g^{*}, \mathbf{F}^{*}_g) - \Pi_{2g}(\mathbf{N}^{a}_{g}, \mathbf{F}^{a}_{g}) - \sum_{j \in \mathcal{J}_{g}^{feas}} (z^*_j(f_j^*) N_{gj}^{*} - z^a_j(f_j^a)N_{gj}^{a})\theta_{fc} 
\end{equation*}
where $z^*_j(f_j^*)$ and $z^a_j(f_j^a)$ denote the observable product characteristics under the optimal and alternative stage-one choices.

Following \citet{yuan2026network}, we use a vector of non-negative instruments $Y$ based on market size quantile bins. Market size is the geometric mean of the two endpoint cities' populations. Each instrument equals one when market size falls within the corresponding bin. These instruments capture variation in observable factors affecting profit changes across markets. We assume that the fixed cost shock difference has zero conditional mean given market size and the characteristics defining each moment group. This restriction implies that its expectation remains zero when multiplied by each bin indicator. We pool observations across quarters and form moments separately by deviation type and, for the redeployment inequalities, by the change in the number of capacity-constrained endpoint airports. Cells with fewer than ten observations are excluded. For each retained instrument $Y_k$:
\begin{equation*}
    \bE[Y_k \cdot \Delta \Pi_{1g} (\mathbf{N}^{*}_{g}, \mathbf{F}^{*}_{g}, \mathbf{N}^{a}_{g}, \mathbf{F}^{a}_{g} ; \theta_{fc})] - \underbrace{\bE [Y_k \cdot \Delta\kappa_g^a]}_{=0} \geq 0
\end{equation*}
Then we construct sample moment inequalities to estimate the fixed cost coefficients $\theta_{fc}$ following \cite{pakes2015moment}:
\begin{equation*}
    -\frac{1}{N^a} \sum_{\mathbf{N}^{*}_{g}, \mathbf{F}^{*}_{g}, \mathbf{N}^{a}_{g}, \mathbf{F}^{a}_{g}} Y_k \cdot \Delta \Pi_{1g} (\mathbf{N}^{*}_{g}, \mathbf{F}^{*}_{g}, \mathbf{N}^{a}_{g}, \mathbf{F}^{a}_{g} ; \theta_{fc}) \leq 0 \hspace{0.5cm} \forall \ k=1,...K 
\end{equation*}
where $N^a$ is the number of feasible alternative product portfolios for airline $g$.

\vspace{-12pt}
\paragraph{Alternative Product Portfolio and Departure Frequency.} Exploring all possible alternative product portfolios and direct departure frequencies is computationally infeasible because the number of $(\mathbf{N}^{a}_{g},\mathbf{F}^{a}_{g})$ combinations grows exponentially with the number of products in airline $g$'s portfolio and the number of feasible departure frequencies for each direct product. Nonetheless, we can construct an outer region covering the identified set by considering a subset of alternative product portfolios and direct departure frequencies. Following \cite{yuan2026network} and \cite{bontemps2023price}, we consider local network deviations: direct-service exit and frequency changes, feasible redeployment to alternative routes, entry into unserved markets, and connecting-flight entry, exit, or connector replacement.

\vspace{-12pt}
\paragraph{Moment Inequalities with Connecting Flights.} Because a deviation on a single direct route may affect multiple markets due to impacts on connecting flights, we must account for these effects. Let $\mM^a$ collect all markets affected by deviation $a$, and let $\pi_{2gm}^*$ and $\pi_{2gm}^a$ denote the airline's expected variable profits across all its products in market $m$ before and after the deviation. Define $\Delta z_g^a$ as observed minus alternative fixed cost characteristics, summed over all affected itineraries, and $\Delta\kappa_g^a$ analogously for fixed cost shocks. The revealed-preference inequality is
\begin{equation}\label{eq:moment inequality}
    \sum_{m\in\mM^a}\left(\pi_{2gm}^*-\pi_{2gm}^a\right)-\Delta z_g^a\theta_{fc}-\Delta\kappa_g^a\geq0.
\end{equation}
This expression covers exit, entry, frequency adjustment, and redeployment. Products outside $\mM^a$ cancel, but surviving products in an affected market remain in the profit comparison because their equilibrium prices and demand can change. Thus neither the direct product's own profit alone nor the profit in its endpoint market alone generally measures the gain from a deviation.

\vspace{-12pt}
\paragraph{Inference.} Following \cite{cox2023simple}, we construct the following conditional likelihood ratio test statistic to conduct inference on the fixed cost parameters $\theta_{fc}$:
\begin{equation} \label{eq:clr}
    CLR(\theta_{fc}) = N\min_{\mu \leq 0} (\hat{m}(\theta_{fc}) - \mu)' \widehat{\Sigma}(\theta_{fc})^{-1} (\hat{m}(\theta_{fc}) - \mu)
\end{equation}
where $\hat{m}(\theta_{fc})$ is the vector of sample moment inequalities, $\widehat{\Sigma}(\theta_{fc})$ consistently estimates the covariance matrix of $\sqrt{N}\hat{m}(\theta_{fc})$, and $\mu$ minimizes the quadratic form subject to $\mu\leq0$. We evaluate the quadratic form on the covariance matrix's non-redundant support. The critical value $\chi_{\hat r,1-\alpha}^{2}$ is the $(1-\alpha)$ quantile of a chi-squared distribution, where $\alpha$ is the significance level and $\hat r$ is the rank of the active constraints in the covariance-standardized system. The confidence region is $CR_{1-\alpha}=\{\theta_{fc}\in\bR^{d_{\theta_{fc}}}:CLR(\theta_{fc})\leq\chi_{\hat r,1-\alpha}^{2}\}$.

\vspace{-12pt}
\section{Estimation Results} \label{sec:estimation_results}

\noindent This section presents and interprets the estimation results for the demand, marginal cost, and fixed cost components of our model.

\vspace{-12pt}
\subsection{Demand and Marginal Cost Estimation}

\noindent As discussed in Section \ref{sec:estimation}, demand parameters are estimated by applying GMM to equation (\ref{eq:demand}). Marginal costs are recovered from the pricing first-order conditions, and the parameters in equation (\ref{eq:marginal_cost}) are estimated by least squares using the same exogenous characteristics and fixed effects as in demand. Table \ref{tab:estimation_results} reports the demand and marginal cost estimation results. The heteroskedasticity-robust Kleibergen--Paap rk Wald F-statistic \citep{kleibergen2006generalized} is 12.19. The nesting parameter $\lambda$ is 0.893, indicating modest similarity of products within a market relative to the outside option.

\begin{table}[!htbp]
\centering
\caption{Demand and Marginal Cost Estimation Results}
\label{tab:estimation_results}
\begin{threeparttable}
\begin{tabular}{@{}lcccc@{}}
\toprule
& \multicolumn{2}{c}{\textbf{Demand}} & \multicolumn{2}{c}{\textbf{Marginal Cost}} \\
\cmidrule(lr){2-3} \cmidrule(lr){4-5}
\textbf{Variable} & \textbf{Coef.} & \textbf{SE} & \textbf{Coef.} & \textbf{SE} \\
\midrule
Constant & $-4.316$ & $(0.534)$ & \phantom{$-$}$0.495$ & $(0.082)$ \\
Fare (\$100) & $-3.561$ & $(0.603)$ &  &  \\
Log Frequency & \phantom{$-$}$1.054$ & $(0.035)$ & \phantom{$-$}$0.054$ & $(0.004)$ \\
Distance & $-0.035$ & $(0.068)$ & \phantom{$-$}$0.019$ & $(0.018)$ \\
Distance$^2$ & \phantom{$-$}$0.181$ & $(0.029)$ & \phantom{$-$}$0.038$ & $(0.005)$ \\
Nesting Parameter & \phantom{$-$}$0.893$ & $(0.037)$ &  &  \\
\bottomrule
\end{tabular}
\footnotesize
\textbf{Note:} Airline-quarter, major airport, airline-hub, and top-50-city fixed effects, non-stop and high-speed rail indicators, and endpoint-type controls and their airline interactions are included but not reported. Standard errors in parentheses are heteroskedasticity-robust for demand and conventional least-squares for marginal cost, conditional on recovered costs. Distance is in 1,000 km; marginal cost is in \$100.
\end{threeparttable}
\end{table}

On average, consumers are willing to pay about \$29.6 for a one-unit increase in the log of daily product frequency, reflecting the high value passengers place on schedule convenience. This is lower than the estimate in \cite{yuan2026network} (\$106.8). The marginal willingness to pay for distance evaluated at the sample average itinerary distance of 1,482 km is \$14.1 per additional 1,000 km, which is lower than the estimate in \cite{yuan2026network} (\$144.34) at the average U.S. domestic flight distance of 1,910 km. This difference reflects many factors including shorter average trip lengths and greater availability of alternative transport modes (e.g., high-speed rail) in Europe.

The average own-price elasticity derived from our parameter estimates is -3.40, which is more elastic than the aggregate price elasticity of -3.13 reported in \cite{marra2024market} for the French market. Compared to estimates from the US market, our estimate is slightly less elastic than the estimate reported in \cite{bontemps2023price} (-3.78) and more elastic than the estimate in \cite{berry2010tracing} (-2.01). Moderately higher elasticities in Europe are consistent with higher European availability of close substitutes, both between airlines in the same market and across alternative modes of transport.

\begin{figure}[htbp]
    \centering
    \includegraphics[width=0.8\textwidth]{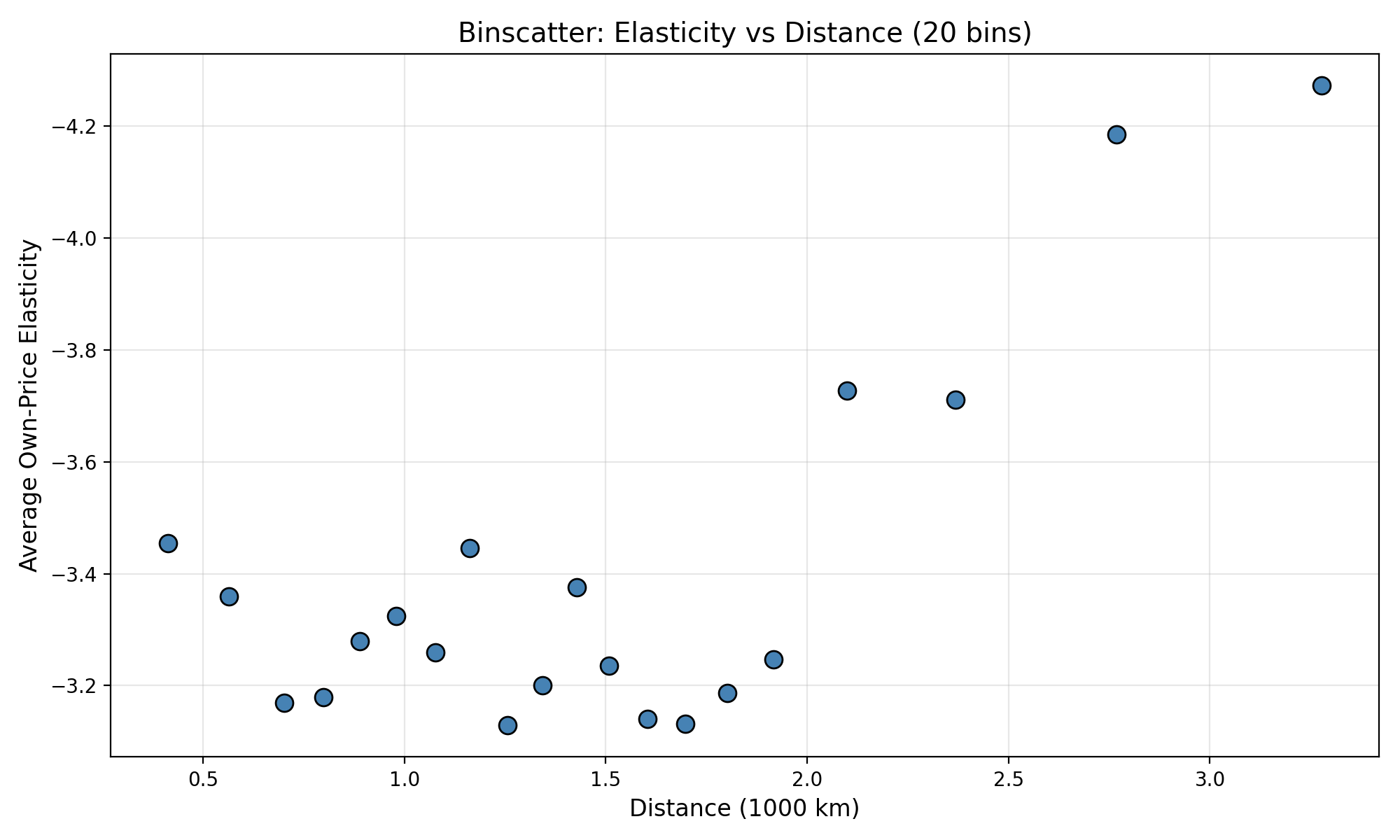}
    \caption{Price Elasticity by Distance}
    \label{fig:elasticity_binscatter}
\end{figure}

Figure \ref{fig:elasticity_binscatter} plots a binscatter of the average price elasticities against itinerary distance. We divide observations into 20 equal-sized bins and plot the average elasticity within each bin. Elasticities range from approximately -3.1 to -3.5 below 2,000 km and become more negative for the longest itineraries, reaching approximately -4.3. The relationship is not monotone over shorter distances. For longer itineraries, the greater price sensitivity indicates that passengers remain responsive even when surface transport is a less convenient substitute. Note that the figure does not hold other market characteristics constant and so reflects variation in elasticities across markets with different prices, shares, and mixes of direct and connecting products.

Table \ref{tab:summary_quarter} presents average price, marginal cost, and variable profit by quarter. The average marginal cost per passenger is \$64.3, about 32\% lower than the estimate in \cite{yuan2026network} (\$95). Part of this difference reflects shorter itineraries and the substantial presence of low-cost carriers in our sample. Given the average itinerary distance of 1,482 km, the per-mile cost is \$0.070. This figure is comparable to \cite{yuan2026network}'s estimate of \$0.08 and \cite{berry2010tracing}'s \$0.06 for U.S. domestic flights. Marginal costs and prices are higher in the second and third quarters than in the first and fourth, consistent with seasonal variation in the composition of services and demand.

\begin{table}[!htbp]
\centering
\caption{Average Price, Marginal Cost, and Variable Profit by Quarter}
\label{tab:summary_quarter}
\begin{threeparttable}
\begin{tabular}{lccc}
\toprule
Quarter & Avg Price & Avg MC & Avg Variable Profit \\
\midrule
Q1 & 86.56 & 59.35 & 0.69 \\
Q2 & 98.42 & 71.24 & 0.75 \\
Q3 & 92.59 & 65.43 & 0.74 \\
Q4 & 86.88 & 59.71 & 0.67 \\
\midrule
Overall & 91.44 & 64.26 & 0.71 \\
\bottomrule
\end{tabular}
\footnotesize
\textbf{Note:} Price and marginal cost are in USD per passenger. Variable profit is in million USD per product-quarter. Entries are unweighted product-quarter means.
\end{threeparttable}
\end{table}

The average price--marginal-cost spread is \$27.2 and the average variable profit per product-quarter is \$0.71 million. Variable profit is revenue less per-passenger marginal costs, before deducting aircraft operating costs and other fixed network costs. We therefore turn to fixed cost estimation before assessing the profitability of the observed network.

\vspace{-12pt}
\subsection{Fixed Cost Estimation}\label{section:fixed_cost_estimation_bounds}

\noindent To evaluate alternative networks, we must impute frequencies for connecting itineraries absent from the data. Following \citet{yuan2026network}, we estimate $\log f_j=a_g+\gamma(\log f_{ah}+\log f_{hb})+u_j$, where $f_j$ is the connecting-flight frequency, $f_{ah}$ and $f_{hb}$ are its two direct-leg frequencies through connecting airport $h$, $a_g$ is an airline fixed effect, and $u_j$ is a regression error. The least squares regression has an $R^2$ of 0.596 and a slope coefficient $\widehat\gamma=0.347$. We convert predictions to frequency levels using the mean exponentiated residual as a correction. For observed connecting itineraries, we use the estimated elasticity to update frequencies when either direct leg changes.

We estimate bounds on the fixed cost parameters using the inequalities in (\ref{eq:moment inequality}). To compute expected profits, we randomly draw $N=36$ pairs of $(\xi,\omega)$ from the empirical joint distribution, with the 2.5th and 97.5th percentiles as the lower and upper bounds of the respective marginals. Drawing the shocks jointly preserves their empirical dependence. The same product draws are used when evaluating alternative frequencies and networks to ensure differences in draws do not drive differences in computed expected profits. As discussed in Section \ref{sec:estimation}, the instruments $Y$ are indicator functions where $Y_k=1$ if a market's population falls within the $k$-th quantile bin. Following \citet{yuan2026network}, we examine progressively finer population-bin instrument sets and report the last nonempty confidence region before further refinement yields an empty region. This data-dependent choice balances informativeness against the feasibility of the estimated restrictions. However, nominal coverage for a fixed instrument set does not automatically extend to a set selected using the same data. We therefore describe the reported region as a nominal 95\% confidence region for the selected specification.

Table \ref{tab:table7} reports the results for seven bins, retaining all bin indicators and using \$5,000 grid increments for each coefficient. The table reports, for each element of the vector $\theta_{fc}$, the projection of the estimated 95\% confidence region in that dimension.

\begin{table}[!htbp]
\centering
\caption{Fixed Cost Estimation}
\label{tab:table7}
\begin{threeparttable}
\footnotesize
\begin{tabular}{@{}cccccccccc@{}}
\toprule
\multicolumn{2}{c}{\textbf{Constant}} & \multicolumn{2}{c}{\textbf{Freq. $\times$ Dist.}} & \multicolumn{2}{c}{\textbf{Log Market Size}} & \multicolumn{2}{c}{\makecell{\textbf{Slot Controlled}\\\textbf{Airport}}} & \multicolumn{2}{c}{\makecell{\textbf{Indirect}\\\textbf{Flight}}}\\
\cmidrule(lr){1-2}\cmidrule(lr){3-4}\cmidrule(lr){5-6}\cmidrule(lr){7-8}\cmidrule(lr){9-10}
\textbf{Lower} & \textbf{Upper} & \textbf{Lower} & \textbf{Upper} & \textbf{Lower} & \textbf{Upper} & \textbf{Lower} & \textbf{Upper} & \textbf{Lower} & \textbf{Upper}\\
\midrule
$1$ & $9$ & $16$ & $18$ & $30.5$ & $37$ & $-41.5$ & $-30$ & $4$ & $4.5$\\
\bottomrule
\end{tabular}
\footnotesize
\textbf{Note:} For each parameter, the table reports the projection of the estimated 95\% confidence region in that dimension. Instruments are population-bin indicators with seven bins. Quarterly cost is in \$10,000, distance in 1,000 km, population in millions, and frequency in directional departures per day. The first four coefficients apply to direct products; the slot variable counts endpoint airports.
\end{threeparttable}
\end{table}

To compare our results to industry benchmarks, we evaluate parameter vectors in the confidence region, compute the fixed cost for each direct flight in each market, average across products and markets, and plot in Figure \ref{fig:distribution_fc_per_flight_hour} the resulting distribution of average fixed cost per flight hour (assuming an average cruising speed of 860 km/h). Our estimated distribution has a median of \$2,945, a 2.5th percentile of \$2,823, and a 97.5th percentile of \$3,086. These estimates are below the industry reported operating costs per flight hour of \$4,829 for A320 family aircraft and \$4,337 for B737 NG aircraft \citep{eurocontrol_ansperformance_aircraft_operating_costs_2024}.\footnote{Our fixed cost component excludes passenger marginal costs and fixed cost shocks, whereas industry operating costs cover a broader set of expenses.}

The frequency--distance coefficient implies distance-related operating costs of \$1,778--2,000 per 1,000 km flown. At an average cruising speed of 860 km/h, this corresponds to \$1,529--1,720 per flight hour. These estimates are below the approximately \$2,450--2,960 per hour implied by fuel and maintenance expenditures in the FY2019 annual reports of easyJet, Wizz Air, and Ryanair.\footnote{Using \citet{easyjet_annual_report_2019}, \citet{wizz_air_annual_report_2019}, and \citet{ryanair_annual_report_2019}, we divide annual fuel and maintenance expenses, converted to USD at 2019 annual average exchange rates, by sectors flown times average sector length in kilometers, and multiply by 860. The resulting hourly equivalents are approximately \$2,957, \$2,452, and \$2,562, respectively.} The coefficient captures fuel, maintenance, and other operating expenses that increase with flight frequency and distance. The total fixed cost per flight hour additionally reflects the constant, market size, and airport-slot terms. Its distribution in Figure \ref{fig:distribution_fc_per_flight_hour} incorporates these components jointly.

The airport-slot coefficient ranges from $-\$415,000$ to $-\$300,000$ per endpoint airport and product-quarter, reflecting a cost advantage conditional on the other characteristics. The additional fixed operating cost of a connecting flight ranges from \$40,000 to \$45,000 per quarter. This is additional to costs incurred on its direct legs, which are already counted as direct operations rather than charged again to each connecting itinerary.

Table \ref{tab:cost_per_pax} reports the marginal cost recovered from the 
pricing first-order conditions and bounds on total cost per passenger computed by adding the deterministic fixed cost. Low-cost carriers such as Ryanair and easyJet have total cost per passenger that is substantially lower than that of full-service carriers like BA/IAG and Lufthansa. Ryanair's total cost is \$44--45 per passenger and easyJet's is \$46--47, compared with \$124--125 for BA/IAG and \$146--147 for Lufthansa. The differences in marginal costs account for much of this ordering, while the fixed cost component also varies with route networks and passenger volumes. These estimates measure modeled operating costs on intra-European routes.

\begin{table}[!htbp]
\centering
\footnotesize
\caption{Marginal Cost and Total Cost per Passenger by Airline}
\label{tab:cost_per_pax}
\begin{threeparttable}
\begin{tabular}{lccc lccc}
\toprule
\multicolumn{4}{c}{Full-Service Carriers} & \multicolumn{4}{c}{Low-Cost Carriers}\\
\cmidrule(lr){1-4}\cmidrule(lr){5-8}
Airline & MC/Pax & LB & UB & Airline & MC/Pax & LB & UB\\
\midrule
British Airways & 100.3 & 123.9 & 125.1 & Ryanair & 21.4 & 43.6 & 45.3\\
Lufthansa & 117.9 & 145.8 & 147.3 & EasyJet & 28.4 & 46.0 & 46.9\\
Air France-KLM & 100.2 & 124.2 & 126.1 & Wizz Air & 19.2 & 43.6 & 46.3\\
SAS & 79.0 & 103.2 & 105.0 & Jet2 & 41.0 & 71.7 & 74.3\\
Finnair & 93.5 & 122.1 & 128.0 & Norwegian & 64.2 & 78.1 & 80.1\\
Aegean & 136.7 & 174.6 & 181.5 & & & &\\
LOT Polish & 172.4 & 231.2 & 233.3 & & & &\\
\midrule
\multicolumn{4}{c}{Regional Carriers} & & & &\\
\cmidrule(lr){1-4}
Icelandair & 97.3 & 130.0 & 135.2 & & & &\\
Air Europa & 61.1 & 81.4 & 83.4 & & & &\\
\bottomrule
\end{tabular}
\footnotesize
\textbf{Note:} MC = Marginal Cost, Pax = Passenger, LB/UB = Lower/Upper Bound of Total Cost per Passenger. All values in USD. Costs are passenger-weighted and exclude fixed cost shocks. Bounds computed across the confidence region.
\end{threeparttable}
\end{table}

\section{Counterfactual Simulations} \label{sec:counterfactual}

\noindent As discussed in Section \ref{sec:background_ets}, cost pressures from carbon pricing mechanisms and fuel supply constraints will increase both fixed operating costs and per-passenger marginal costs. To quantify how these cost pressures will affect airline network and pricing decisions, we simulate four counterfactual scenarios that increase costs by \$1.0--\$2.5 per kilometer flown. We label these scenarios Low, Medium, High, and Very High and use the abbreviations Low, Med, High, and VH.

To simulate the impact on fixed costs, we increase the fixed-cost coefficient on Frequency $\times$ Distance by 9, 13.5, 18, and 22.5, respectively, in units of \$10,000. To simulate the impact on per-passenger marginal costs we add a per-passenger carbon surcharge to marginal cost, proportional to total itinerary distance. The surcharge equals $\tau\times3.16\times2.5/1{,}000$ dollars per passenger per 1,000 km, where $\tau$ is the carbon price in dollars per tonne, 3.16 is the CO$_2$ emission factor, and 2.5 is marginal fuel use in kg per additional passenger per 1,000 km.\footnote{The emission factor follows \citet{icao_aircraft_engine_emissions_databank} and marginal fuel use follows \citet{steinegger2017fuel}.} At carbon prices of \$100, \$150, \$200, and \$250, the passenger surcharges are \$0.79, \$1.185, \$1.58, and \$1.975, respectively. Aircraft charges apply once per direct leg. Passenger charges cover both legs of connecting itineraries.

Then in each scenario we simulate counterfactual outcomes for each quarter in our sample using a local-deviation algorithm that iteratively updates airline entry, frequency, and pricing decisions until convergence. Following \citet{yuan2026network}, the equilibrium concept is a restricted subgame perfect equilibrium in which airlines are restricted to local deviations. Details of the algorithm are provided in Appendix~\ref{sec:counterfactual_equilibrium}.

\vspace{-12pt}
\subsection{Carbon Regulation Simulation Results} \label{sec:counterfactual_results}

\noindent We classify markets as short-haul if market distance is at most 1,500 km and as medium- and long-haul otherwise. Market distance is the average observed direct-product distance within the city-pair, using endpoint distance when no direct product is observed.

We simulate counterfactuals separately for four quarters and four carbon price scenarios. For each scenario we compute impacts on passenger numbers, prices, carbon emissions,  consumer surplus, profits, carbon tax revenue, and increases in social welfare due to reduced carbon emissions. Carbon emissions reductions are valued at \$215 per tonne of $CO_2$. Table \ref{tab:cf_results} reports the aggregate welfare effects, and Table \ref{tab:airline_breakdown} decomposes outcomes by airline type.

\begin{table}[!htbp]
\centering
\footnotesize
\setlength{\tabcolsep}{4pt}
\caption{Counterfactual Results}
\label{tab:cf_results}
\begin{tabular}{@{}lrrrrr@{}}
\toprule
 & \multicolumn{1}{c}{\textbf{Base}} & \multicolumn{1}{c}{\textbf{Low}} & \multicolumn{1}{c}{\textbf{Med}} & \multicolumn{1}{c}{\textbf{High}} & \multicolumn{1}{c@{}}{\textbf{VH}}\\
\midrule
\multicolumn{6}{@{}l}{\textbf{All Markets}}\\
Consumer Surplus & 9.55\phantom{ (0.00)} & -13.50 (0.49)\% & -21.97 (0.40)\% & -27.80 (0.28)\% & -33.94 (0.30)\%\\
Profit & 3.50 (0.04) & -1.42 (1.15)\% & -16.60 (1.16)\% & -30.01 (1.12)\% & -41.78 (1.07)\%\\
Carbon Revenue &  & \phantom{-}1.94 (0.02)\phantom{\%} & \phantom{-}2.36 (0.02)\phantom{\%} & \phantom{-}2.70 (0.02)\phantom{\%} & \phantom{-}2.91 (0.02)\phantom{\%}\\
Social Value of CO$_2$ &  & \phantom{-}2.66 (0.04)\phantom{\%} & \phantom{-}3.30 (0.02)\phantom{\%} & \phantom{-}3.69 (0.01)\phantom{\%} & \phantom{-}4.02 (0.02)\phantom{\%}\\
Welfare Gain &  & \phantom{-}3.26 (0.04)\phantom{\%} & \phantom{-}2.98 (0.05)\phantom{\%} & \phantom{-}2.68 (0.05)\phantom{\%} & \phantom{-}2.22 (0.05)\phantom{\%}\\
Passengers & 331.90\phantom{ (0.00)} & -14.28 (0.47)\% & -22.84 (0.39)\% & -28.73 (0.27)\% & -34.89 (0.29)\%\\
\midrule
\multicolumn{6}{@{}l}{\textbf{Short-haul Markets}}\\
Consumer Surplus & 7.05\phantom{ (0.00)} & \phantom{-}11.73 (0.63)\% & \phantom{-}2.25 (0.53)\% & -4.64 (0.39)\% & -12.37 (0.40)\%\\
Profit & 2.65 (0.02) & \phantom{-}8.90 (1.11)\% & -8.01 (1.12)\% & -23.41 (1.07)\% & -37.06 (1.02)\%\\
Carbon Revenue &  & \phantom{-}1.75 (0.02)\phantom{\%} & \phantom{-}2.18 (0.02)\phantom{\%} & \phantom{-}2.52 (0.01)\phantom{\%} & \phantom{-}2.73 (0.02)\phantom{\%}\\
Social Value of CO$_2$ &  & \phantom{-}0.65 (0.04)\phantom{\%} & \phantom{-}1.17 (0.02)\phantom{\%} & \phantom{-}1.50 (0.01)\phantom{\%} & \phantom{-}1.79 (0.01)\phantom{\%}\\
Welfare Gain &  & \phantom{-}3.46 (0.03)\phantom{\%} & \phantom{-}3.29 (0.04)\phantom{\%} & \phantom{-}3.07 (0.04)\phantom{\%} & \phantom{-}2.67 (0.04)\phantom{\%}\\
Passengers & 244.64\phantom{ (0.00)} & \phantom{-}10.79 (0.61)\% & \phantom{-}1.19 (0.52)\% & -5.79 (0.37)\% & -13.55 (0.39)\%\\
\midrule
\multicolumn{6}{@{}l}{\textbf{Medium- and long-haul Markets}}\\
Consumer Surplus & 2.50\phantom{ (0.00)} & -84.73 (0.42)\% & -90.34 (0.39)\% & -93.16 (0.34)\% & -94.82 (0.27)\%\\
Profit & 0.85 (0.03) & -33.73 (1.60)\% & -43.52 (1.65)\% & -50.69 (1.61)\% & -56.59 (1.56)\%\\
Carbon Revenue &  & \phantom{-}0.19 (0.01)\phantom{\%} & \phantom{-}0.19 (0.01)\phantom{\%} & \phantom{-}0.18 (0.01)\phantom{\%} & \phantom{-}0.17 (0.01)\phantom{\%}\\
Social Value of CO$_2$ &  & \phantom{-}2.01 (0.01)\phantom{\%} & \phantom{-}2.13 (0.01)\phantom{\%} & \phantom{-}2.19 (0.01)\phantom{\%} & \phantom{-}2.23 (0.01)\phantom{\%}\\
Welfare Gain &  & -0.20 (0.01)\phantom{\%} & -0.31 (0.02)\phantom{\%} & -0.39 (0.02)\phantom{\%} & -0.45 (0.02)\phantom{\%}\\
Passengers & 87.26\phantom{ (0.00)} & -84.60 (0.42)\% & -90.23 (0.40)\% & -93.06 (0.34)\% & -94.75 (0.28)\%\\
\bottomrule
\end{tabular}
\par\smallskip
\begin{minipage}{\linewidth}\footnotesize
\textbf{Note:} Entries are means across 50 parameter and shock draws, with standard deviations in parentheses. Update orders are averaged within each draw. Monetary levels are in USD billions and passenger levels in millions. Scenario entries for consumer surplus, profit, and passengers are percentage changes from baseline. Profit deducts fixed costs, including fixed-cost shocks. Short-haul markets are at most 1,500 km. Welfare Gain $=$ $\Delta$Profit $+$ $\Delta$CS $+$ Carbon Revenue $+$ Social Value of CO$_2$.
\end{minipage}
\end{table}

\begin{table}[!htbp]
\centering
\footnotesize
\setlength{\tabcolsep}{4pt}
\caption{Counterfactual Results by Airline Type}
\label{tab:airline_breakdown}
\begin{tabular}{@{}lrrrrr@{}}
\toprule
\textbf{Metric} & \multicolumn{1}{c}{\textbf{Base}} & \multicolumn{1}{c}{\textbf{Low}} & \multicolumn{1}{c}{\textbf{Med}} & \multicolumn{1}{c}{\textbf{High}} & \multicolumn{1}{c@{}}{\textbf{VH}}\\
\midrule
\multicolumn{6}{@{}l}{\textbf{All}}\\
Profit & 3.50 (0.04) & -1.42 (1.15)\% & -16.60 (1.16)\% & -30.01 (1.12)\% & -41.78 (1.07)\%\\
Fare & 92.42 (0.00) & \phantom{-}0.15 (0.28)\% & -2.79 (0.24)\% & -4.85 (0.22)\% & -7.47 (0.25)\%\\
Direct departures & 29.79 (0.00) & -14.05 (0.78)\% & -24.22 (0.60)\% & -30.68 (0.43)\% & -37.21 (0.41)\%\\
Distance Flown & 31,627.42 (0.00) & -43.80 (0.64)\% & -54.50 (0.41)\% & -61.00 (0.25)\% & -66.43 (0.26)\%\\
FC/Pax & 17.03 (0.11) & -9.76 (1.09)\% & -5.87 (1.09)\% & \phantom{-}0.11 (1.03)\% & \phantom{-}5.83 (1.06)\%\\
Cost/Pax & 81.87 (0.11) & -1.77 (0.48)\% & -4.19 (0.41)\% & -5.24 (0.34)\% & -7.07 (0.37)\%\\
\midrule
\multicolumn{6}{@{}l}{\textbf{Full-Service}}\\
Profit & 1.70 (0.01) & -13.51 (1.66)\% & -27.01 (1.79)\% & -39.00 (1.61)\% & -51.66 (1.50)\%\\
Fare & 132.82 (0.00) & -1.31 (0.14)\% & -0.49 (0.27)\% & \phantom{-}0.07 (0.31)\% & \phantom{-}0.58 (0.32)\%\\
Direct departures & 17.43 (0.00) & -23.08 (1.34)\% & -40.43 (1.03)\% & -51.37 (0.72)\% & -62.34 (0.69)\%\\
Distance Flown & 16,450.43 (0.00) & -46.56 (1.07)\% & -62.41 (0.65)\% & -71.02 (0.39)\% & -77.90 (0.41)\%\\
FC/Pax & 15.84 (0.10) & -1.42 (2.17)\% & -7.32 (2.59)\% & -8.39 (2.44)\% & -9.33 (2.81)\%\\
Cost/Pax & 121.12 (0.10) & -1.70 (0.34)\% & -1.38 (0.43)\% & -0.75 (0.45)\% & -0.21 (0.47)\%\\
\midrule
\multicolumn{6}{@{}l}{\textbf{Low-Cost}}\\
Profit & 1.73 (0.03) & \phantom{-}10.11 (1.56)\% & -6.64 (1.35)\% & -21.36 (1.24)\% & -32.18 (1.12)\%\\
Fare & 59.52 (0.00) & \phantom{-}5.49 (0.16)\% & \phantom{-}6.07 (0.16)\% & \phantom{-}6.94 (0.15)\% & \phantom{-}8.29 (0.17)\%\\
Direct departures & 11.76 (0.00) & -0.02 (0.00)\% & -0.03 (0.00)\% & -0.05 (0.01)\% & -0.06 (0.01)\%\\
Distance Flown & 14,355.58 (0.00) & -39.56 (0.26)\% & -44.76 (0.24)\% & -49.02 (0.23)\% & -52.88 (0.22)\%\\
FC/Pax & 18.00 (0.15) & -15.12 (0.82)\% & -5.83 (0.82)\% & \phantom{-}2.58 (0.87)\% & \phantom{-}9.15 (0.92)\%\\
Cost/Pax & 49.88 (0.15) & \phantom{-}1.53 (0.36)\% & \phantom{-}5.46 (0.38)\% & \phantom{-}9.44 (0.37)\% & \phantom{-}13.23 (0.40)\%\\
\midrule
\multicolumn{6}{@{}l}{\textbf{Regional}}\\
Profit & 0.07 (0.00) & \phantom{-}7.66 (3.49)\% & -9.90 (2.87)\% & -25.31 (2.97)\% & -39.05 (2.56)\%\\
Fare & 97.25 (0.00) & -14.50 (0.57)\% & -15.00 (0.54)\% & -14.97 (0.56)\% & -13.88 (0.69)\%\\
Direct departures & 0.60 (0.00) & -26.79 (0.26)\% & -27.36 (0.20)\% & -30.11 (1.21)\% & -35.34 (1.78)\%\\
Distance Flown & 821.41 (0.00) & -62.65 (0.81)\% & -66.41 (0.59)\% & -69.90 (0.67)\% & -73.59 (0.80)\%\\
FC/Pax & 16.77 (0.27) & -28.29 (2.64)\% & -16.67 (2.30)\% & -5.92 (2.82)\% & \phantom{-}3.21 (2.78)\%\\
Cost/Pax & 86.89 (0.27) & -21.83 (0.86)\% & -20.17 (0.83)\% & -18.22 (0.78)\% & -15.40 (0.73)\%\\
\bottomrule
\end{tabular}
\par\smallskip
\begin{minipage}{\linewidth}\footnotesize
\textbf{Note:} Means and standard deviations are computed as in Table~\ref{tab:cf_results}. Base reports levels. Scenario columns report percentage changes from baseline. Profit is in USD billions. Fare and costs are in USD per passenger. Direct departures and distance flown are in 100,000 departures and 100,000 km, respectively. FC/Pax includes fixed-cost shocks. Cost/Pax adds marginal costs, including the carbon surcharge. Per-passenger values divide annual totals by passengers.
\end{minipage}
\end{table}

\vspace{-12pt}
\paragraph{Baseline Validation.} The baseline generates annual net profit of \$3.50 billion (s.d.\ 0.04), after deducting deterministic fixed costs and realized fixed-cost shocks from expected variable profit of \$9.15 billion. Relative to revenue of \$30.67 billion, the profit margin is 11.4\%, within the range of the 7.3\%, 10.2\%, and 13.2\% operating margins calculated from easyJet, IAG, and Ryanair's 2019 accounts.\footnote{Authors' calculations divide operating profit by revenue. Accounts cover entire networks and ancillary activities. Model net profit deducts modeled operating costs, but is not corporate profit after financing costs and income taxes.} Consumer surplus is \$9.55 billion and total passenger numbers are 331.90 million. Expected passenger traffic and variable profit are 4.70\% and 4.98\% below their observed counterparts.

Short-haul markets contribute \$7.05 billion of consumer surplus, \$2.65 billion of profit (s.d.\ 0.02), and 244.64 million passengers. Medium- and long-haul markets contribute \$2.50 billion, \$0.85 billion (s.d.\ 0.03), and 87.26 million passengers, respectively. The longer-distance segment thus accounts for about one-quarter of traffic.

\vspace{-12pt}
\paragraph{Aggregate Welfare Effects.} Under Low, consumer surplus falls by 13.5\% and profits by 1.4\%, so consumers bear a larger proportional loss. However, profit losses grow faster with the tax. Under VH they reach 41.8\%, compared with 33.9\% for consumer surplus. The reduction in passenger volumes rises from 14.3\% to 34.9\%.

Carbon revenue and avoided emissions more than offset these private losses. Revenue rises from \$1.94 billion under Low to \$2.91 billion under VH. The social value of avoided emissions rises from \$2.66 billion to \$4.02 billion. Welfare gains decline from \$3.26 billion (s.d.\ 0.04) to \$2.22 billion (s.d.\ 0.05). Low therefore maximizes welfare among the scenarios. Additional emissions benefits at higher rates grow more slowly than private-surplus losses.

Higher distance-related costs induce exit from longer routes and redeployment toward shorter city pairs. Marginal-cost increases put upward pressure on prices for a given network, while exit and frequency reductions diminish product variety and schedule convenience. Passengers on affected routes substitute toward other itineraries or the outside option. Capacity redeployment can improve short-haul services, and changes in product composition mean average fares need not rise. On the supply side, shorter flights lower operating costs, offsetting much of the variable-profit loss under Low. At higher taxes, fewer profitable redeployment opportunities remain.

\vspace{-12pt}
\paragraph{Asymmetric Effects by Market Segment.} Aircraft and passenger charges both scale with distance, exposing longer routes more strongly. Under Low, medium- and long-haul consumer surplus falls by 84.7\% (s.d.\ 0.4), whereas short-haul surplus rises by 11.7\% (s.d.\ 0.6). Under VH, longer-distance surplus falls by 94.8\% (s.d.\ 0.3). Passenger losses rise from 84.6\% under Low to 94.7\% under VH.

Short-haul consumers benefit from redirected capacity at lower rates. Surplus remains 2.3\% above baseline under Med, before falling by 4.6\% under High and 12.4\% under VH. Profits similarly rise by 8.9\% (s.d.\ 1.1) under Low, compared with a 33.7\% loss (s.d.\ 1.6) in longer markets. Under VH, both segments lose profit, but the longer-distance decline is larger, at 56.6\% versus 37.1\%.

Environmental benefits are also uneven. Longer markets provide \$2.01 billion of the \$2.66 billion social value of avoided emissions under Low. Even including emissions benefits and tax revenue, their welfare changes remain negative, from $-\$0.20$ billion under Low to $-\$0.45$ billion under VH. Short-haul welfare gains range from \$3.46 billion to \$2.67 billion. Positive aggregate welfare therefore combines gains in shorter markets with losses on country-pairs dependent on longer routes.

\vspace{-12pt}
\paragraph{Heterogeneous Responses by Airline Type.} Table~\ref{tab:airline_breakdown} distinguishes FSCs, LCCs, and regional carriers (Icelandair and Air Europa).

\textit{Profit impacts.} Under Low, FSC profits fall by 13.5\% (s.d.\ 1.7), whereas LCC and regional profits rise by 10.1\% (s.d.\ 1.6) and 7.7\% (s.d.\ 3.5). Surviving services can have lower operating costs or more favorable competitive conditions despite falling aggregate traffic. The regional gain is small in dollars. Baseline profit is \$0.07 billion, compared with \$1.70 billion for FSCs and \$1.73 billion for LCCs. All three lose at higher rates. Under VH, losses are 51.7\% (s.d.\ 1.5), 32.2\% (s.d.\ 1.1), and 39.1\% (s.d.\ 2.6), respectively.

\textit{Pricing responses.} LCC fare increases range from 5.5\% under Low to 8.3\% under VH. FSC fares fall by 1.3\% under Low, remain close to baseline under High, and rise by 0.6\% under VH. Regional fares fall by approximately 14--15\%. The surcharge is proportionally larger relative to LCC marginal costs. These passenger-weighted averages also reflect product exit and changing shares. Removing longer, expensive services can lower average fares despite higher route-level costs.

\textit{Network adjustments.} LCCs mainly reallocate departures from longer to shorter routes. Their total frequency falls by only 0.02\% under Low and 0.06\% under VH, but distance falls by 39.6\% and 52.9\%. FSC frequency falls by 23.1\% and 62.3\%, alongside distance reductions of 46.6\% and 77.9\%. Regional frequency falls by 26.8\% and 35.3\%, and distance by 62.7\% and 73.6\%. All groups shorten the average direct departure, but only LCCs largely preserve departure totals.

\textit{Cost structure.} LCC fixed cost per passenger falls by 15.1\% under Low as shorter routes reduce operating costs, then rises by 9.2\% under VH. FSC fixed cost per passenger remains below baseline, by 1.4--9.3\%, whereas regional costs initially fall by 28.3\% before rising by 3.2\% under VH. Both passenger volumes and fixed costs change, so their relative movements determine these ratios. Total cost per passenger rises by 1.5--13.2\% for LCCs but falls by 0.2--1.7\% for FSCs and 15.4--21.8\% for regional carriers. These averages include fixed cost shocks and reflect network recomposition, helping explain why traffic reductions need not translate proportionately into lower profits.

\vspace{-12pt}
\paragraph{Summary.} Carbon revenue and avoided emissions outweigh private-surplus losses under our tax and valuation assumptions. Adjustment is highly asymmetric across carrier types and market segments. Shorter markets gain welfare, longer markets lose, and FSCs face the largest proportional profit loss at the highest tax. Reallocation toward shorter direct routes is central, with LCCs preserving departures while other carriers also contract frequency.

\vspace{-12pt}
\subsection{Geographic Distribution of Welfare Effects} \label{sec:geographic}

\noindent Figures~\ref{fig:heatmap_welfare_bl}--\ref{fig:heatmap_profit_uh} depict the geographic distribution of net welfare, consumer surplus, and profit across European country-pairs. For each outcome, the figures show both baseline values (Figures \ref{fig:heatmap_welfare_bl}, \ref{fig:heatmap_cs_bl}, and \ref{fig:heatmap_profit_bl}) and impacts of the Low and Very High carbon tax scenarios (remaining figures). We omit figures for the intermediate scenarios. We use the market-distance measure and short-haul classification defined in Section~\ref{sec:counterfactual_results}.

The figures display results for the top 15 country groups ranked by baseline welfare, accounting for 82.0\% of total welfare. Each figure displays one observation per country-pair in the lower triangle, with domestic markets on the diagonal. The tax scenario figures report changes relative to the no-tax baseline. The tax scenario titles display the aggregate impacts. All monetary values are in billions of dollars.

The baseline figures confirm that airline market welfare, consumer surplus, and profits are concentrated in a relatively small set of high-traffic country-pairs. Spain--UK contributes approximately \$0.97 billion of baseline welfare, followed by domestic Spain at \$0.59 billion and France--Spain at \$0.43 billion. This concentration means that even modest proportional changes on the busiest corridors can have large effects on aggregate surplus. It also highlights the distinction between a country's total activity and outcomes in particular bilateral markets.

The tax scenario panels show how this concentration changes as airlines redeploy capacity. Country-pairs dependent on longer itineraries can lose passenger service and consumer surplus even when the social value of avoided emissions is substantial. Other pairs benefit from capacity redirected toward shorter routes. Positive aggregate welfare gains therefore coexist with losses on particular country-pairs.

\vspace{-12pt}
\section{Conclusion} \label{sec:conclusion}

\noindent This paper estimates a two-stage model of airline competition with endogenous route entry, departure frequency, and pricing to quantify the impact of carbon regulation on the European airline industry. Our counterfactual simulations reveal that the effects of carbon pricing under the EU ETS are highly asymmetric. Full-service carriers reduce total direct departure frequency by 23--62\% and regional carriers by 27--35\%, while low-cost carriers adjust primarily by reallocating departures toward shorter direct routes. Consumer surplus declines by up to 34\% overall, with medium- and long-haul markets bearing the brunt, while short-haul markets experience positive net welfare gains as airlines reallocate capacity toward shorter routes. Airline profits decline by 1--42\%, but carbon tax revenue of \$1.9--2.9 billion together with a social value of avoided CO\textsubscript{2} emissions of \$2.7--4.0 billion more than offset these private-surplus losses. These results demonstrate that carbon regulation can achieve meaningful environmental gains in European aviation, though at significant and unevenly distributed cost.

\clearpage
\setlength{\bibsep}{0pt}
\setcitestyle{authoryear,round}
\bibliographystyle{abbrvnat}
\bibliography{reference.bib}

\appendix
\clearpage
\section{Counterfactual Simulation Algorithm} \label{sec:counterfactual_equilibrium}
\vspace{-12pt}
The counterfactual simulation algorithm is as follows. First we specify feasible local route and frequency deviations and draw 50 parameter vectors and fixed-cost shocks. We hold these draws fixed throughout the analysis. Then for each draw, we simulate restricted subgame perfect equilibria using the algorithm below.

\vspace{-12pt}
\paragraph{Feasible Deviations.}
Route deviations are restricted to local network deviations that change a single route at a time. Feasible deviations include: exit, frequency reduction or increase, or redeployment. Deviations must satisfy the consideration set, slot-capacity, and fleet capacity constraints in Section~\ref{sec:model}. Frequency increases require released departure capacity. Connecting services can be added, canceled, or assigned a different connecting airport if both direct legs remain active. Profit comparisons include all products in affected markets, so an unprofitable direct product may support profitable connections.

To account for aircraft capacity constraints, we add a smooth capacity penalty to marginal cost when implied passengers per flight exceed 200. The penalty takes the form $\text{penalty}_j=2.28\ln(1+\exp(L_j-200))$ dollars per passenger, where $L_j$ is the implied load. The penalty is negligible sufficiently below the threshold and rises approximately linearly above it. It discourages concentrating passenger traffic on an unrealistically small number of flights, but it is not a hard capacity limit.

\begingroup
\setlength{\emergencystretch}{1em}
Frequency deviations, measured in directional departures per day, are restricted to market-specific bins drawn from the set
\begin{equation*}
    \left\{
    \left(0,\frac{1}{7}\right],\left(\frac{1}{7},\frac{2}{7}\right],
    \left(\frac{2}{7},\frac{4}{7}\right],
    \left(\frac{4}{7},2\right],
    \left(2,4\right],\left(4,6\right],\left(6,8\right],
    \left(8,10\right],10+
    \right\}.
\end{equation*}
\endgroup
To ensure feasible frequency deviations are empirically relevant, for each market cell defined by market size and distance, we select frequency bins that account for at least 90\% of observed direct-departure frequency choices. This gives four to six bins per cell. The distance cells use the market-distance measure defined in Section~\ref{sec:counterfactual_results}, with short-haul markets at or below 1,500 km and medium- and long-haul markets above 1,500 km. Frequencies for indirect flights are determined by the frequencies of their component legs through the relationship in Section \ref{section:fixed_cost_estimation_bounds}.

 \vspace{-12pt}
\paragraph{Parameters and shocks.}
For the counterfactuals, we add two assumptions. First, we assume the fixed-cost parameters are uniformly distributed over the 1,147 accepted parameter vectors in the estimated confidence region. Second, we assume that fixed-cost shocks are drawn from a truncated normal distribution, truncated to ensure that at the baseline, the observed product portfolio is more profitable than feasible deviations. The truncated normal shocks, before truncation, have mean zero and airline-specific variances equal to 5\% of the variance of observed direct products' deterministic fixed costs \citep{wollmann2018trucks}.

For routes and frequencies that are in the consideration set but are not feasible in the baseline (i.e., routes or frequencies that are blocked by slot or frequency constraints), we draw fixed-cost shocks from the unconditional normal distribution. For each fixed-cost parameter vector draw, we draw one complete set of fixed-cost shocks for all airlines and all feasible routes and frequencies. The fixed-cost shock draws are held fixed throughout all counterfactual simulations.

 \vspace{-12pt}
\paragraph{Simulation Algorithm.} 
We sequentially evaluate local deviations for all airlines in all markets. We follow \citet{wollmann2018trucks, yuan2026network,bontemps2023price} and consider multiple orders of evaluation to explore the potential range of network outcomes. We consider three orders of network evaluation. The first order is based on variable profits. For each quarter, we rank markets and airlines within a market based on variable profits. If an airline has multiple direct routes in the same market, we rank those routes based on their variable profits. Then, we begin with the market with the highest total variable profits. For each airline in that market, starting from the airline with the highest variable profits, we evaluate all feasible deviations for each direct product. If an airline finds a profitable deviation, it deviates accordingly. After all airlines in the market have been evaluated and potential deviations executed, we proceed to the next market in the variable profit ranking and repeat the process.

After the direct-product sweep, we evaluate feasible connecting-product deviations on the resulting direct operating network. Each accepted action updates the affected markets, and we re-solve their price competition before assessing the next relevant profit comparison. At the end of each iteration, we update the rankings of markets, airlines, and routes based on the new variable profits.

The second order reverses the ranking, starting with the markets and products with the lowest total variable profits. The third order randomizes the evaluation order. Although direct decisions are anchored in currently active products, their redeployment targets need not be currently served by the airline. We therefore do not assume that every unserved market must remain unprofitable after the network changes.

For each order, we repeat the process until no airline can find a profitable supported deviation or the sequence revisits a previous network. In the latter case, we report average aggregate statistics across the cycle, rather than treating any one state as a stable network.\footnote{Cycle averages summarize the algorithm's outcomes. They are not established mixed-strategy equilibria. Runs reaching the iteration limit without convergence or a completed cycle are excluded.}

To report results, we aggregate outcomes to the annual level, average across completed network evaluation sequences within each parameter and shock draw, and then average across the 50 draws. The reported standard deviations describe variation across these draws after averaging evaluation orders.

\clearpage
\section{Supplementary Figures} \label{app:heatmaps} \label{app:supplementary}
\par\bigskip
\noindent\begin{minipage}{\textwidth}
    \centering
    \captionsetup{font=small,hypcap=false}
    \includegraphics[width=0.65\textwidth]{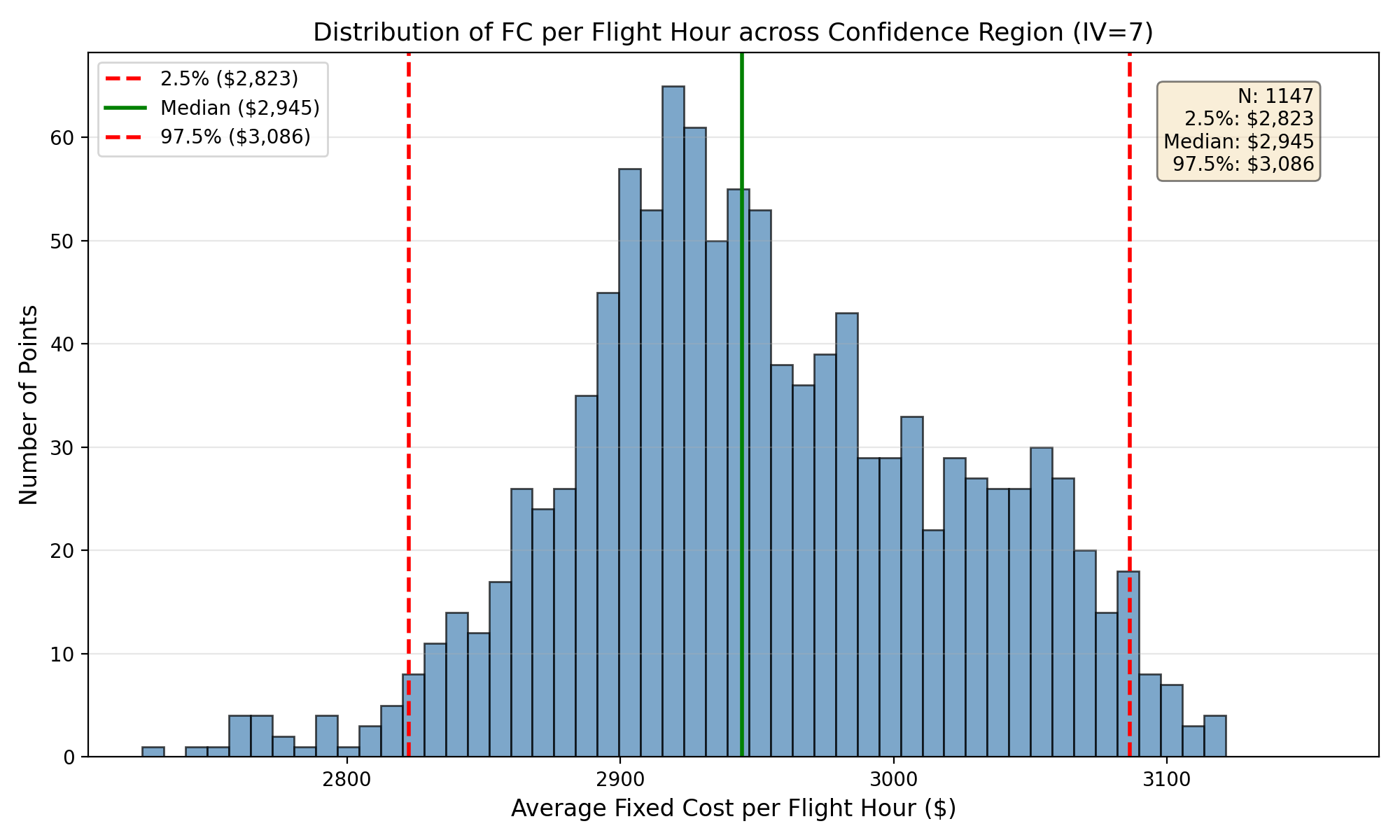}
    \captionof{figure}{Distribution of Fixed Cost per Flight Hour}
    \label{fig:distribution_fc_per_flight_hour}
\end{minipage}\par

\begingroup
\captionsetup{font={footnotesize,stretch=1},hypcap=false}

\newcommand{\geoheatmap}[3]{%
  \noindent
  \begin{minipage}[t]{0.49\textwidth}
    \centering
    \includegraphics[width=\linewidth]{#1}
    \captionof{figure}{#2}
    \label{#3}
  \end{minipage}
}

\geoheatmap{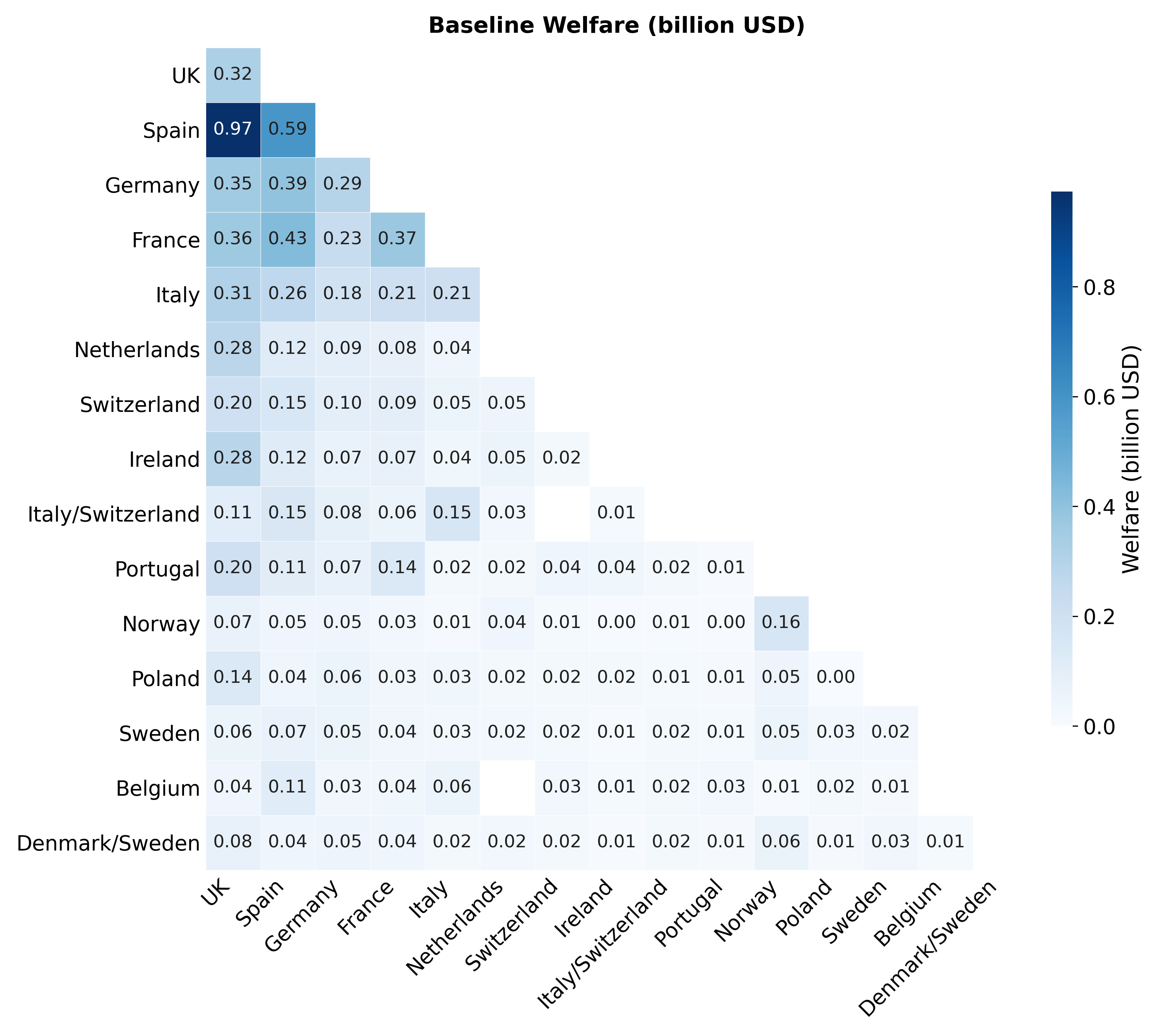}{Welfare by country-pair: baseline.}{fig:heatmap_welfare_bl}\hfill
\geoheatmap{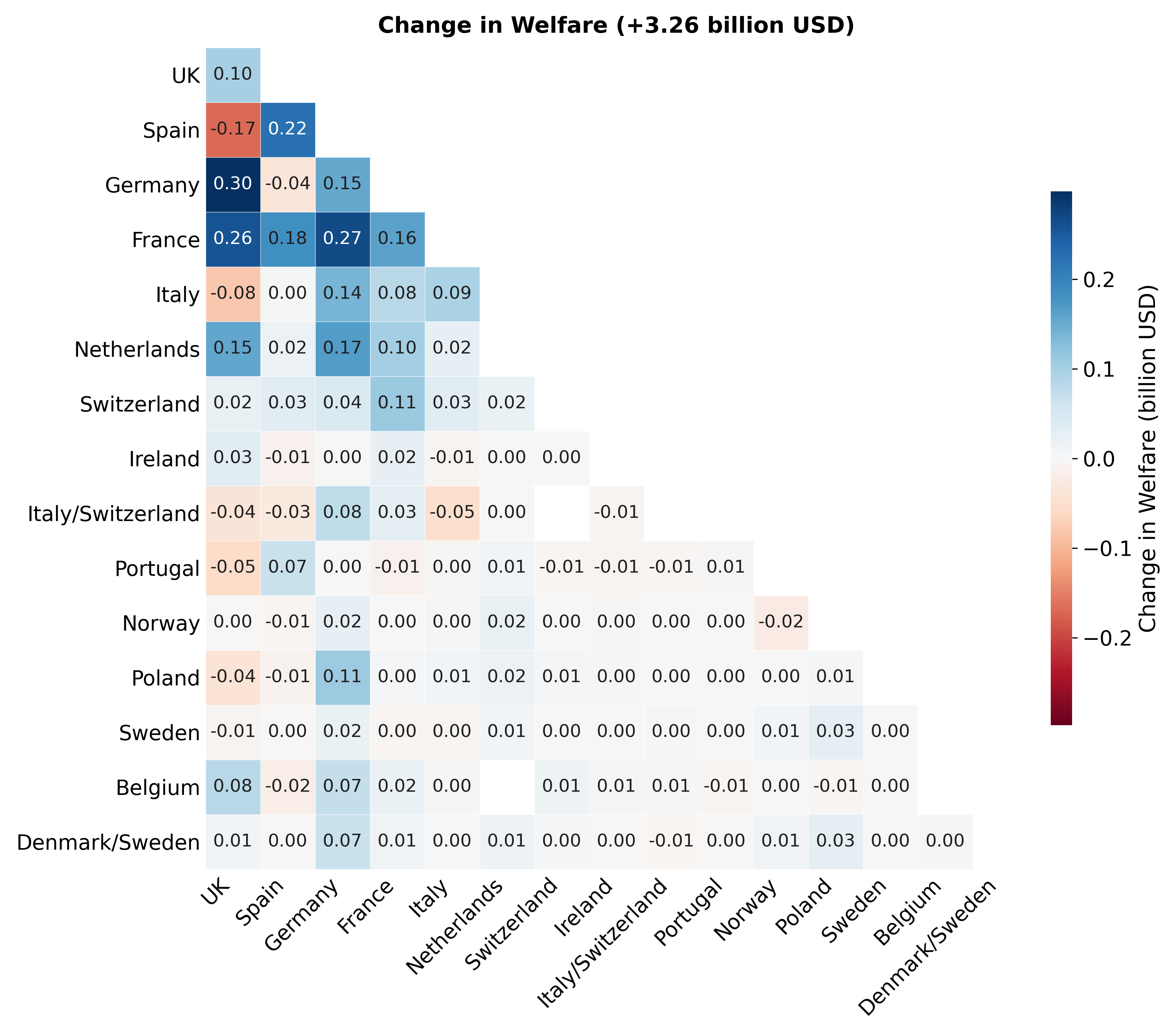}{Welfare change by country-pair: Low scenario.}{fig:heatmap_welfare_low}
\par\medskip
\geoheatmap{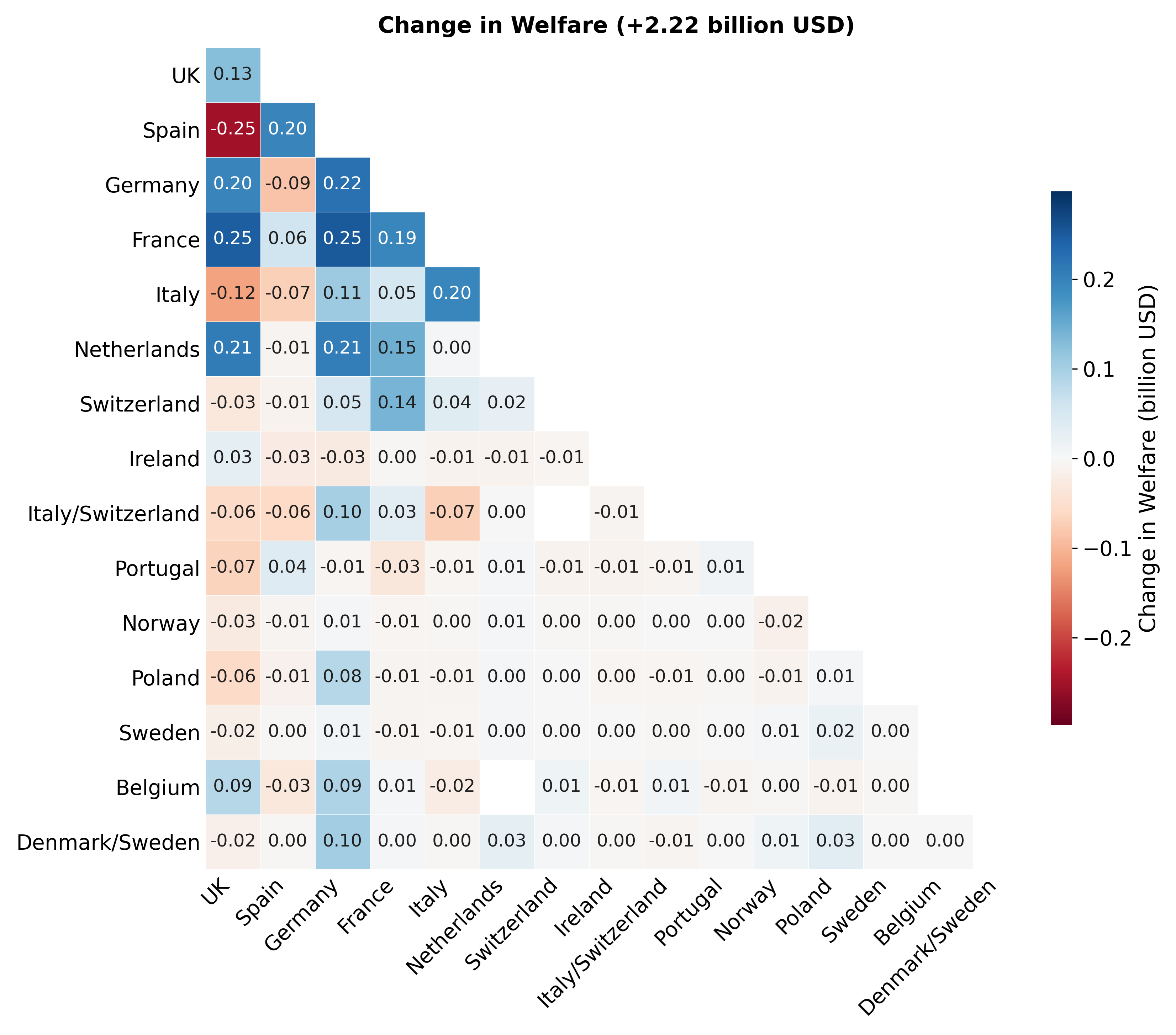}{Welfare change by country-pair: Very High scenario.}{fig:heatmap_welfare_uh}\hfill
\geoheatmap{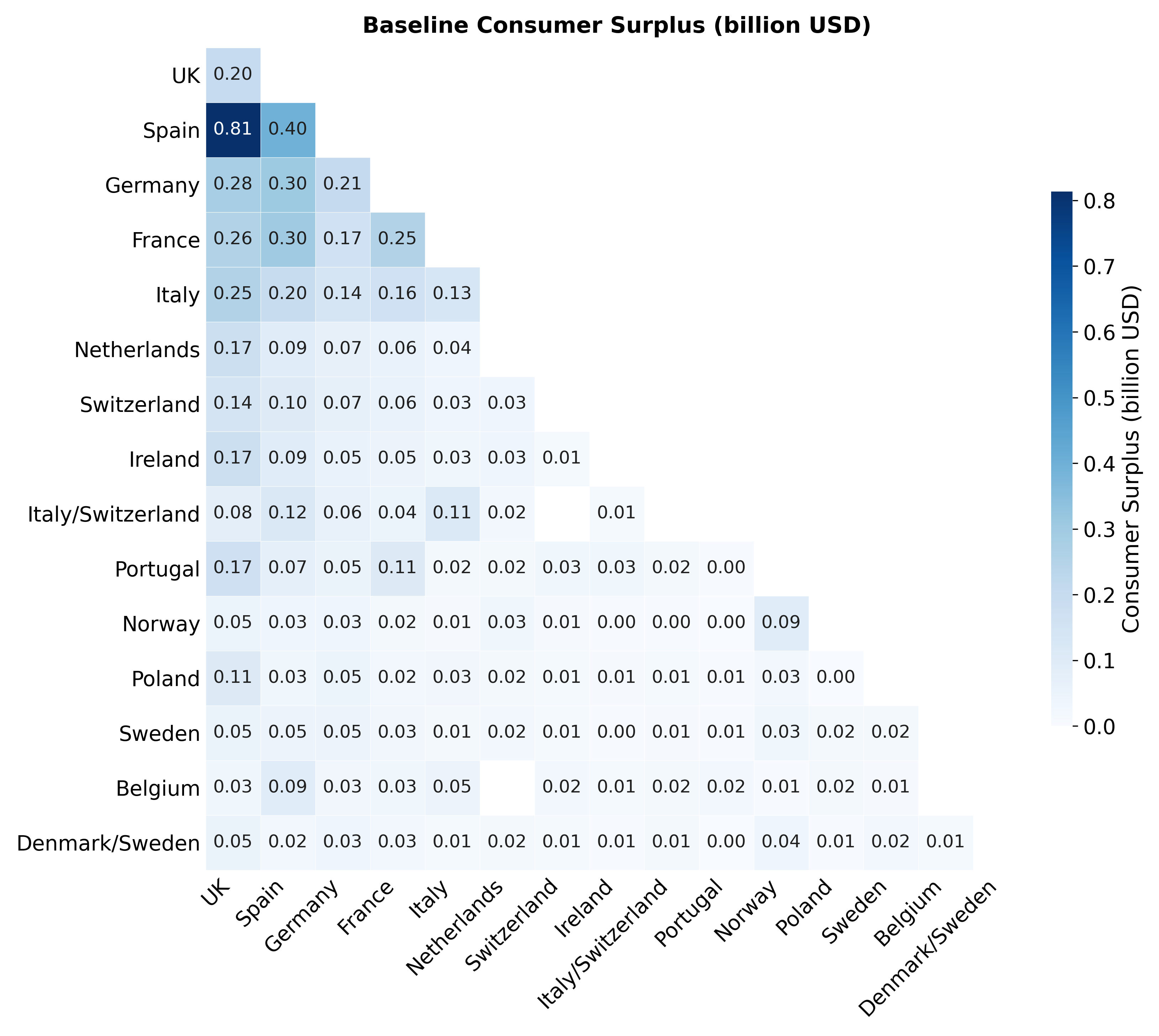}{Consumer surplus by country-pair: baseline.}{fig:heatmap_cs_bl}
\par\medskip
\geoheatmap{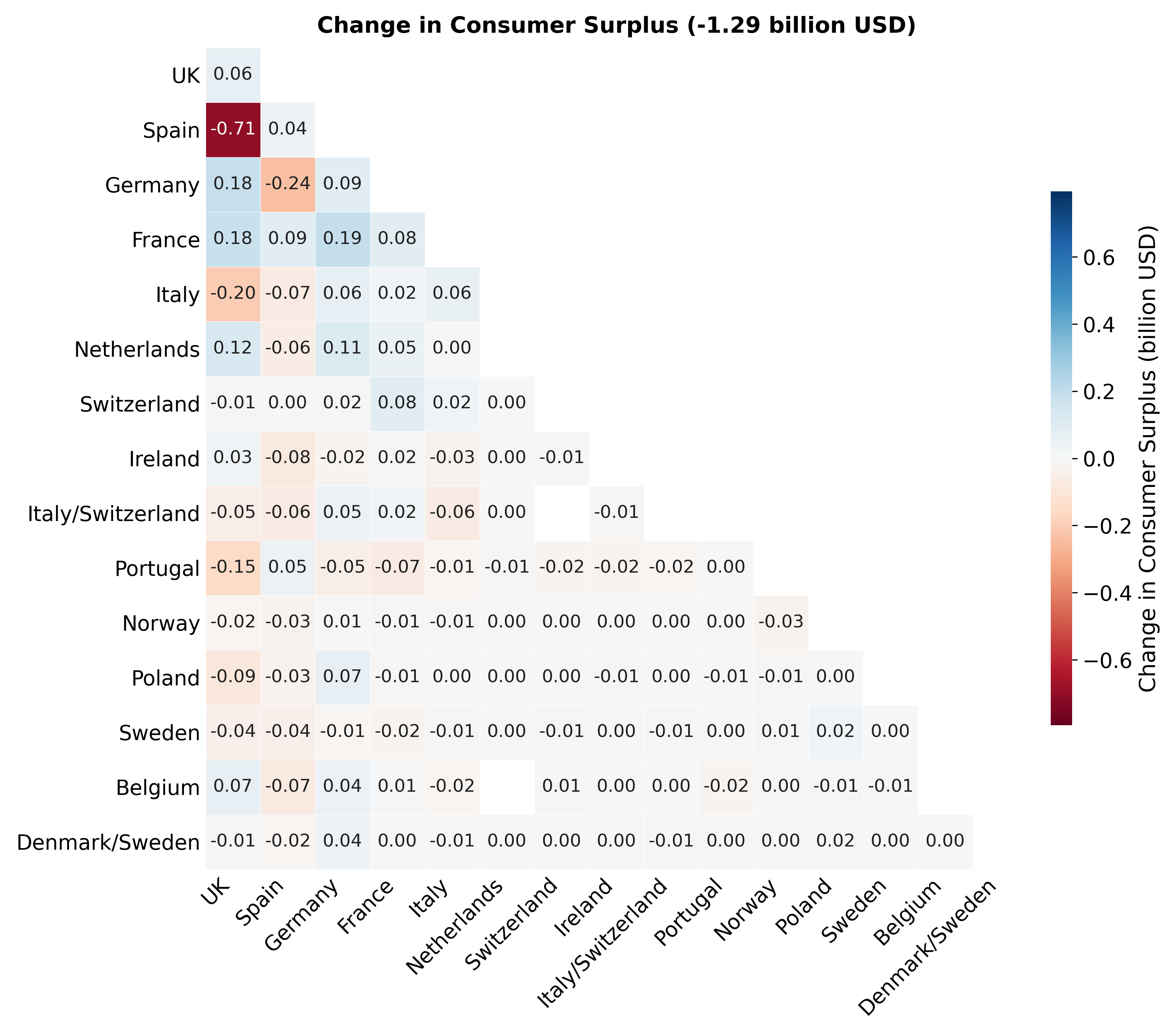}{Consumer surplus change by country-pair: Low scenario.}{fig:heatmap_cs_low}\hfill
\geoheatmap{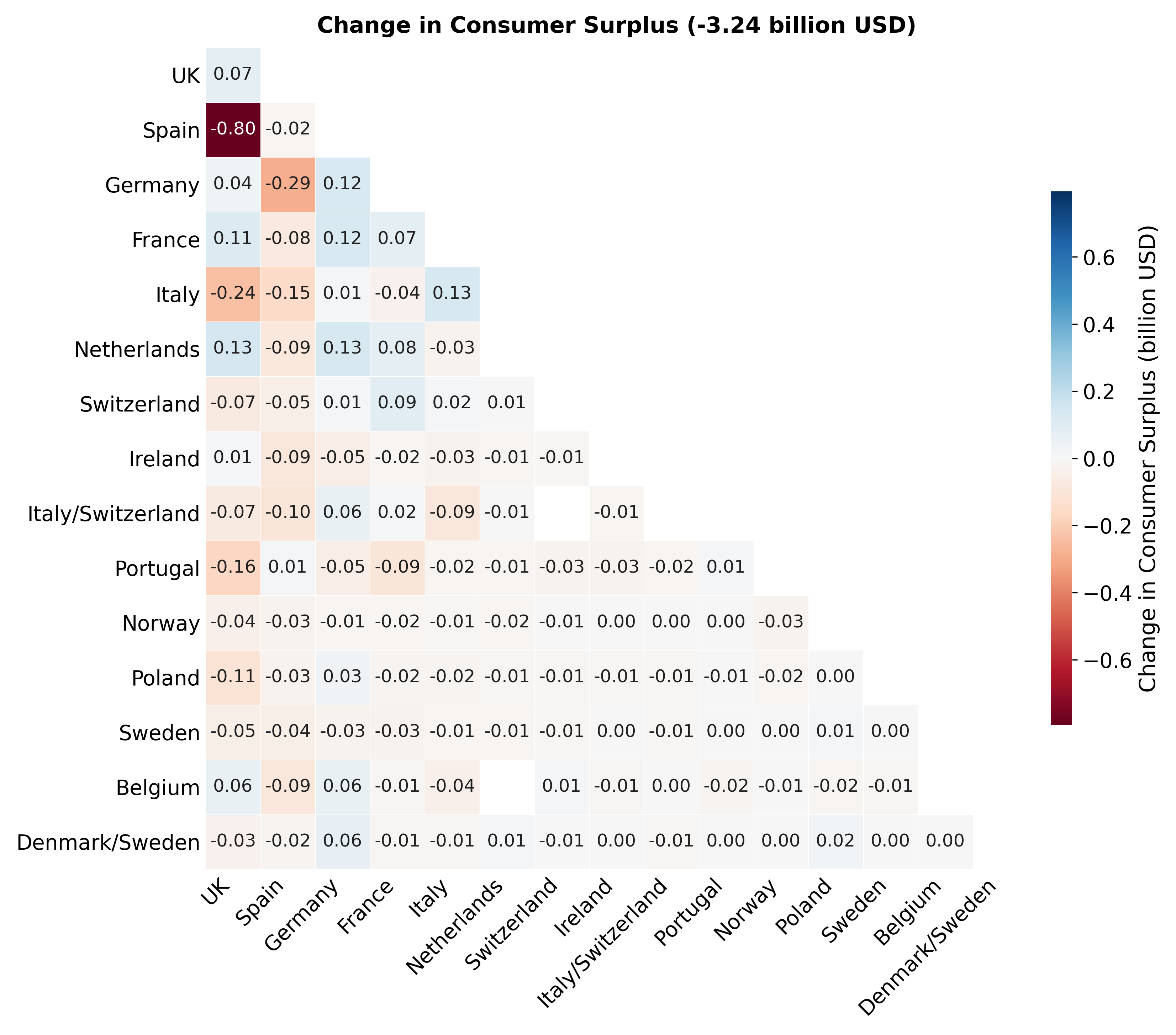}{Consumer surplus change by country-pair: Very High scenario.}{fig:heatmap_cs_uh}
\par\medskip
\geoheatmap{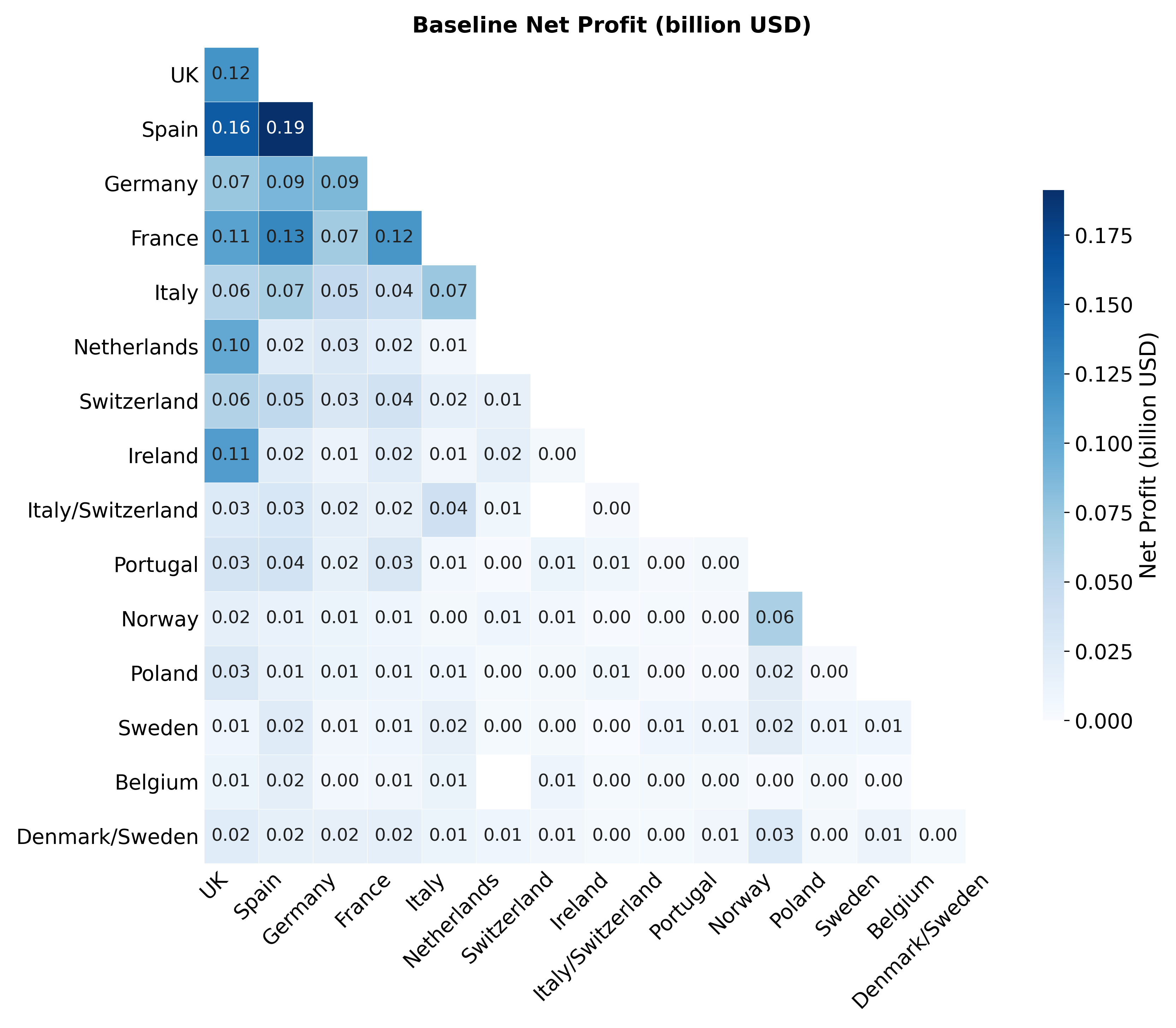}{Net profit by country-pair: baseline.}{fig:heatmap_profit_bl}\hfill
\geoheatmap{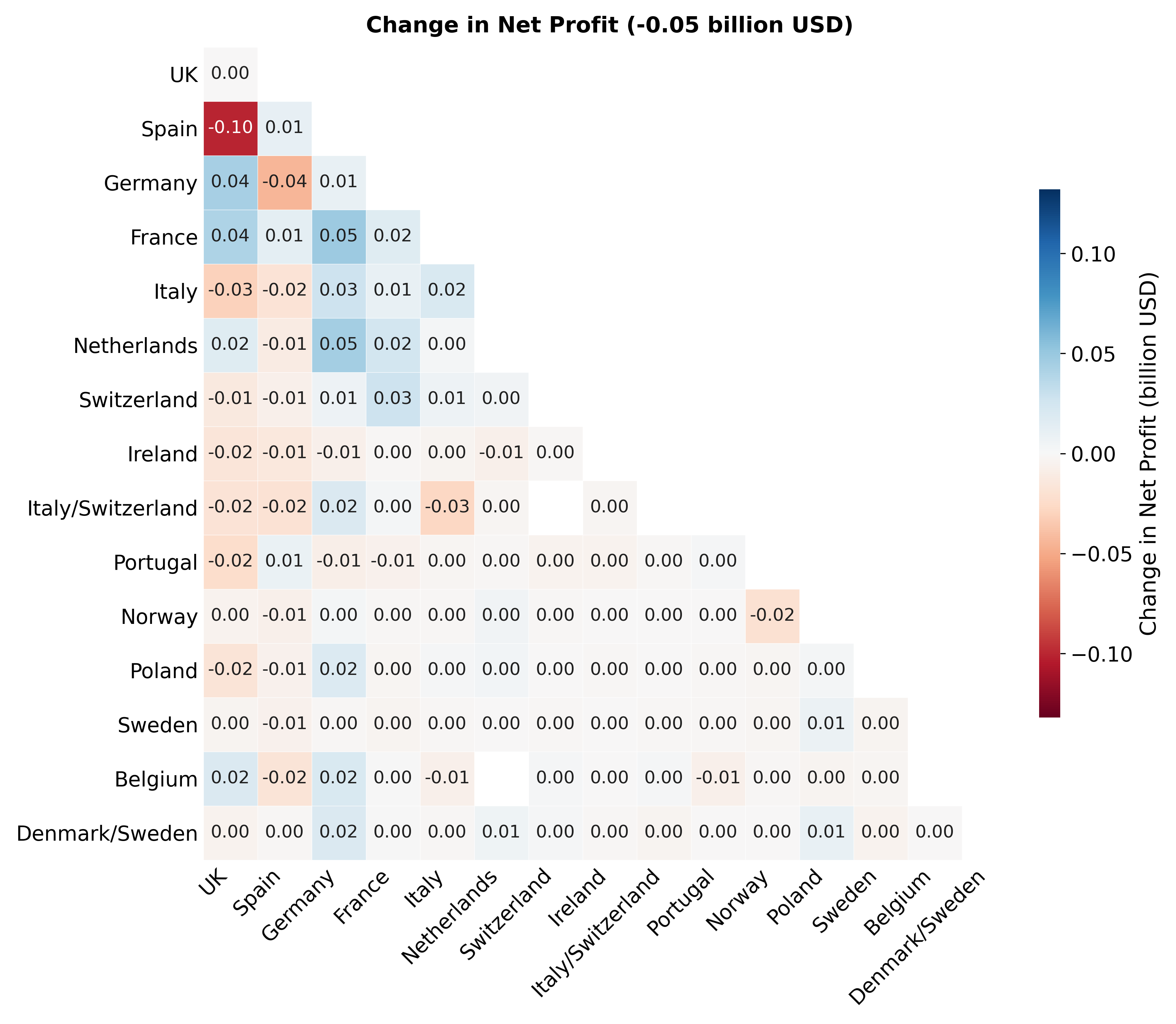}{Net profit change by country-pair: Low scenario.}{fig:heatmap_profit_low}
\par\medskip
\noindent\hfill
\geoheatmap{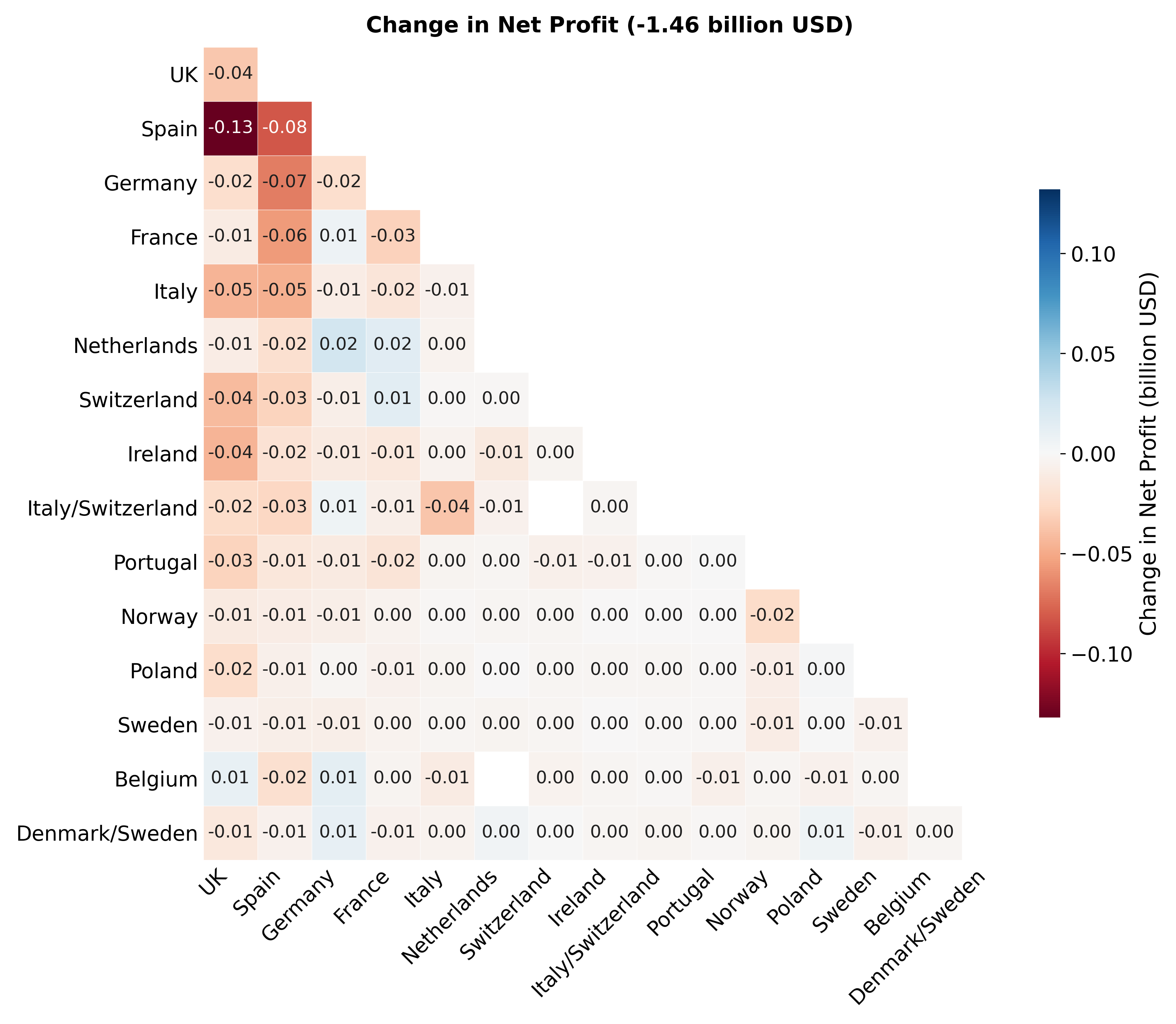}{Net profit change by country-pair: Very High scenario.}{fig:heatmap_profit_uh}
\hfill\mbox{}

\endgroup

\end{document}